\documentclass[journal]{IEEEtran}
\usepackage{cite}
\usepackage{bigints}
\usepackage{amsmath}
\usepackage{array}
\usepackage[pdftex]{graphicx}
\usepackage{xcolor}
\usepackage{hyperref}
\title{Numerically stable eigenmode extraction in 3D periodic metamaterials}

\author{%
D. Tihon \textit{Student Member, IEEE},
\and V. Sozio, \textit{Member, IEEE},
\and N.A. Ozdemir, 
\and M. Albani, \textit{Senior Member, IEEE}
and C. Craeye \textit{Senior Member, IEEE}\thanks{Some parts of this work was supported by the Research Networking Program NEWFOCUS of the European Science Foundation "New Frontiers in mm/sub-mm waves integrated dielectric focusing systems".
} \thanks{D. Tihon, N. Ozdemir and C. Craeye are with the Universit\'{e} Catholique de Louvain, in the ICTEAM institute, Place du Levant 3, 1348 Louvain-la-Neuve, Belgium (e-mails: denis.tihon@uclouvain.be, nilufer.ozdemir@uclouvain.be, christophe.craeye@uclouvain.be).} \thanks{V. Sozio and M. Albani are with the University of Siena, Dept. of Information Eng. and Mathematics, Via Roma, 56, 53100 Siena, Italy (e-mails: valentina.sozio@virgilio.it, matteo.albani@dii.unisi.it).}}

\begin{document}

\maketitle

\begin{abstract}
A numerical method is presented to compute the eigenmodes supported by three dimensional (3D) metamaterials using the Method of Moments (MoM). The method relies on interstitial equivalent currents between layers. First, a parabolic formulation is presented. Then, we present an iterative technique that can be used to linearize the problem. In this way, all the eigenmodes characterized by their transmission coefficients and equivalent interstitial currents can be found using a simple eigenvalue decomposition of a matrix. 
The accuracy that can be achieved is only limited by the quality of simulation, and we demonstrate that the error introduced when linearizing the problem decreases doubly exponentially with respect to the time devoted to the iterative process. We also draw a mathematical link and distinguish the proposed method from other transfer-matrix based methods available in the literature.
\end{abstract}

\begin{IEEEkeywords}
Numerical modeling, eigenmode analysis, metamaterials.
{\color{blue} \textcopyright 2016 IEEE.  Personal use of this material is permitted.  Permission from IEEE must be obtained for all other uses, in any current or future media, including reprinting/republishing this material for advertising or promotional purposes, creating new collective works, for resale or redistribution to servers or lists, or reuse of any copyrighted component of this work in other works. The published version is available at \url{https://doi.org/10.1109/TAP.2016.2562659}.}
\end{IEEEkeywords}

\section{Introduction}
The numerical analysis of metamaterials can be challenging, given the small size of the inclusions and their subwavelength periodicity, which induces a very strong coupling between neighboring unit-cells. One way to circumvent this limitation is to simplify the complex structure using approximation techniques, such as homogenization \cite{1}. These techniques generally rely on quasi-analytical formulations based on specific hypotheses. However, a deep understanding of the phenomena taking place at the scale of the inclusions is required to properly select the hypotheses. 

One way to characterize numerically the response of an infinite metamaterial to any excitation at a given frequency is to compute its dispersion diagram and the corresponding Bloch modes of the structure at the frequency of interest. Interesting information can be extracted from this diagram \cite{rev2:3}, such as the behaviour of the Poynting vector \cite{rev2:1}, or the presence of photonic band gaps \cite{rev2:2}. However, the process can be numerically complex if the problem is not properly tackled \cite{bookYasumoto}. 
Fast techniques based on plane wave decomposition have been proposed \cite{context1, 3}. In \cite{context5}, a more efficient method based on hybrid plane-wave-integral-equations is proposed in the context of metallic crystals. A method based on root-searching is provided in \cite{context4}. Methods of \cite{4,5} are based on a combination of Transfer (T)-matrix formalism and plane-wave decomposition of the fields, the latter partly compensating for the ill-conditioning of the T-matrix. However, this issue is confirmed in \cite{limitations}, where the authors propose to use the symmetry properties of the eigenvalues and the Rayleigh quotients to relieve the accuracy limitations.

A method to circumvent the issue of T-matrix ill-conditioning has been proposed in \cite{last-min} for the computation of the dispersion diagrams. However, information about the shape of the modes is then lost.

% the order-N method \cite{context2}, or methodologies based on transfer matrix \cite{3,4,5}. Methods based on root searching are also presented in \cite{context3, context4}. A more efficient method, based on hybrid plane-wave-integral-equation is provided in \cite{context5} to compute the band structure of three-dimensional metallic crystals, without resorting to root-searching.

In this paper, we propose a general method to compute the dispersion diagram and the shape of the corresponding eigenmodes of any 3D periodic metamaterial, limiting its study to one unit-cell and using the Poggio-Miller-Chan-Harrington-Wu-Tsai (PMCHWT) formulation applied to the Method of Moments \cite{7}. This formulation is based on imposing the continuity of tangential fields across interfaces using equivalent currents. Imposing also the continuity of the fields across the interface separating two consecutive layers of the metamaterial and imposing the periodicity of the currents between consecutive layers within a complex transmission coefficient, it is possible to determine the eigenmodes inside the structure. In a first instance, the resulting formulation of the problem is non-linear w.r.t. transmission coefficient of the eigenmodes. To solve this problem, we propose an iterative technique applied to the impedance matrix and demonstrate how it linearizes the problem such that all the modes |propagative and evanescent| characterized by a given transverse wave-vector can be computed at once with great accuracy without significantly increasing the complexity of the method. Preliminary results concerning this method have been provided in \cite{2}. 

The presented method can be transposed into the Scattering (S-)matrix formalism, and is therefore related to \cite{4,5}. However, using the iterative technique, the problem can be linearized without resorting to the T-matrix, which is known to be ill-conditioned \cite{8}. Consequently, any field decomposition (i.e. not only plane waves \cite{4,5}) can be used.

 % more classical methods based on transfer or scattering matrix formulations, as have been devised in the context of photonic crystal research \cite{4,5}. {\color{red} + Briefly announce the advantage over other methods}

The remainder of this paper is divided as follows. In Section 2, the eigenvalue problem is introduced and the non-linear formulation is derived. In Section 3, the iterative technique that is applied to the impedance matrix and that linearizes the formulation is presented; its convergence is discussed in Section 4. In Section 5, the formulation is extended to the scattering matrix approach and a comparison with other techniques is established. Finally, in Section 6, numerical results are presented and commented before concluding in Section 7.

\section{Non-Linear Formulation}
\label{sec:non-lin}
We first focus on a single layer of metamaterial, as the one illustrated in Fig. \ref{fig:1layer}. In order to decouple the layer from the outer medium, equivalent currents are placed on the two interfaces separating it from outer space \cite{nil_curr_2, 6}. Using the equivalent currents, the inside of the layer is isolated from the outer source distribution and the inner and outer problems can be treated separately. The 3D metamaterial under study is made of an infinite stack of such layers. The system of coordinates is chosen such that the layer is periodic along the $\hat{x}$ and $\hat{y}$ directions with periods of $d_x$ and $d_y$ respectively, and perpendicular to the $\hat{z}$ direction. 

Such a 2D infinite layer can be analyzed using the Method of Moments (MoM). Using equivalent currents on the interface separating different media, Maxwell's equations are solved for each medium separately using the Green's function formalism. The distribution of the equivalent currents is chosen such that, summing the solutions obtained for the different media composing the layer, the tangential fields are continuous across interfaces (PMCHWT formulation \cite{PMCHWT}). Infinitely periodic layers can be simulated using periodic Green's function \cite{7}.

The equivalent currents on the surface of the inclusion and on the surfaces separating consecutive layers are discretized using Rao-Wilton-Glisson (RWG) \cite{RWG} and rooftop basis functions. The total current distribution $\mathbf{I}(\mathbf{r})$ is described using a vector $\mathbf{x} = [x_1~x_2...]^T$ such that 

\begin{equation}
\mathbf{I}(\mathbf{r}) = \sum_m x_m \mathbf{F}_m(\mathbf{r})
\end{equation}

with $\mathbf{F}_m(\mathbf{r})$ the current distribution associated to the $m^{th}$ basis function. The unknown coefficients $x_m$ are determined imposing the continuity of tangential fields on both sides of the interfaces between different media. The continuity of the fields is tested over testing function, which are generally identical to the basis functions (Galerkin testing). The system of equations to be solved becomes

\begin{equation}
Z \mathbf{x} = -\mathbf{b}
\end{equation}

with $\mathbf{b}$ the vector whose $l^{th}$ entry corresponds to the external fields evaluated on testing function $l$ and $Z$ the impedance matrix, whose $(l,m)^{th}$ entry corresponds to the fields radiated by currents of unit amplitude on basis function $m$ and evaluated on testing function $l$.

%Such 2D infinite layer can be simulated using the Method of Moments (MoM). The solution of the Maxwell's equations inside the whole piecewise homogeneous medium is divided into smaller problems, each one corresponding to one distinct homogeneous piece. The different pieces are isolated one from each other using equivalent currents on their surface. The total problem can be solved finding a combination of solutions of the sub-problems for which the sum of the currents on the surfaces vanish and the boundary conditions are satisfied one the interface between different media.

%Such infinitely periodic layer can be simulated using the Method of Moments. First, the study of the infinite layer can be brought back to the study of a single layer using the periodic Green's function instead of the classical one \cite{7}. The surfaces of the inclusions and the surface separating consecutive layers are discretized using Rao-Wilton-Glisson (RWG) {\color{red}[]} or rooftop {\color{red}[]} basis functions. Using PMCHWT formulation, equivalent currents on these surfaces have to found, such that the tangential fields are continuous over the interfaces. The field continuity is generally tested over the same discretized mesh (Galerkin testing). 

The study of the periodic layer can be simplified by using the periodic Green's function when computing the impedance matrix \cite{7}. Considering a given phase shift of the fields and currents $\varphi_x$ and $\varphi_y$ between consecutive unit-cells in the $\hat{x}$ and $\hat{y}$ directions respectively, the equivalent currents on both sides of the layer and on the surface of the inclusions can be found by solving the following system of equations:

\begin{figure}[h!]
\center
\includegraphics[width = 8cm]{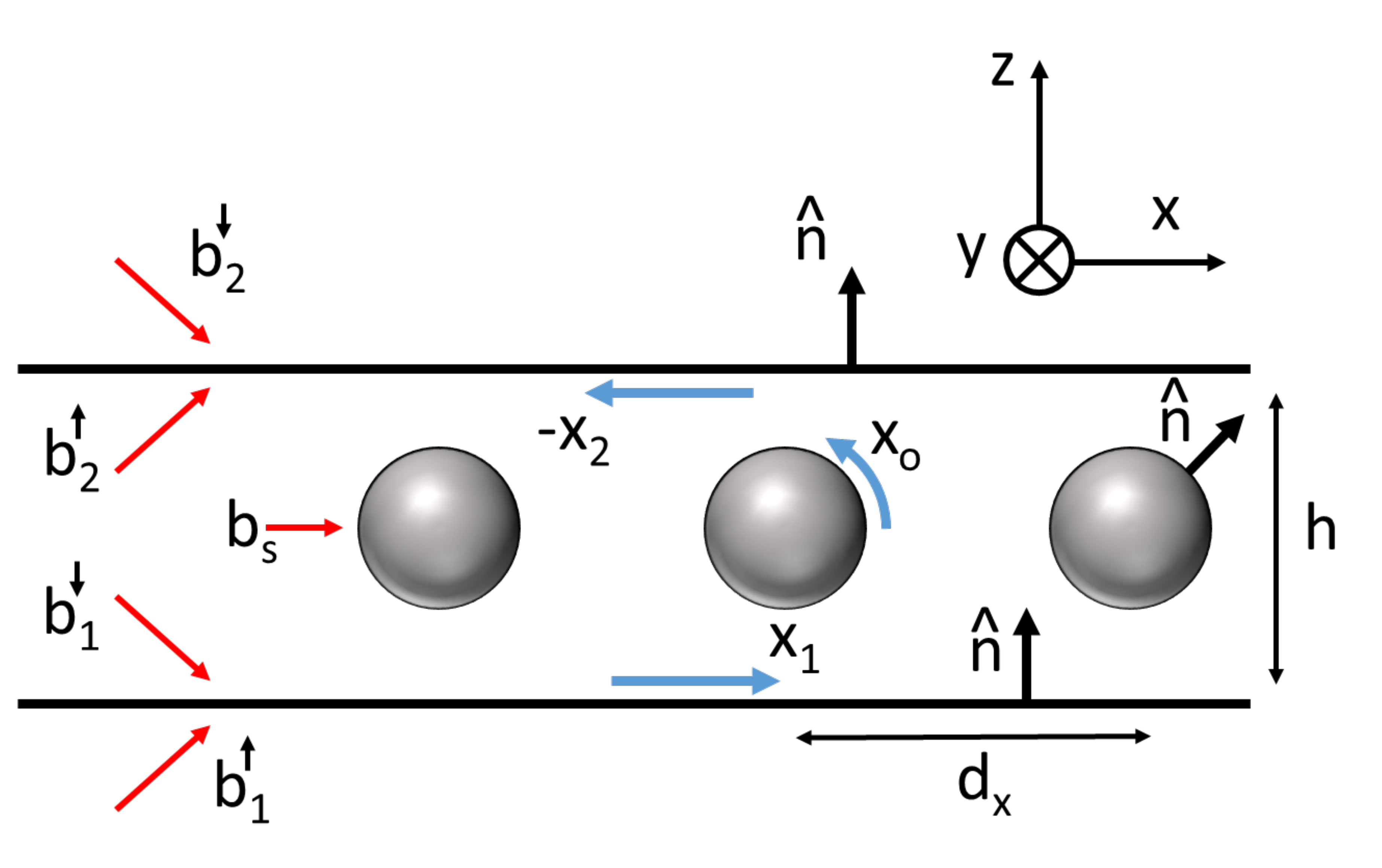}
\caption{One layer of the metamaterial under study. The fields exciting the structure are represented with red arrows, while equivalent currents are represented using blue arrows. The normals of the surfaces are defined as indicated by the $\hat{n}$ arrows.}
\label{fig:1layer}
\end{figure}

\begin{equation}
\label{eq:8-05-1}
\begin{pmatrix}
\tilde{Z}_{11} & \tilde{Z}_{1s} & -\tilde{Z}_{12} \\
\tilde{Z}_{s1} & \tilde{Z}_{ss} & -\tilde{Z}_{s2} \\
-\tilde{Z}_{21} & -\tilde{Z}_{2s} & \tilde{Z}_{22}
\end{pmatrix}
\cdot
\begin{pmatrix}
\mathbf{x}_1 \\
\mathbf{x}_s \\
\mathbf{x}_2
\end{pmatrix}
= -
\begin{pmatrix}
\mathbf{b}_1 \\
\mathbf{b}_s \\
\mathbf{b}_2
\end{pmatrix}
\end{equation}

with $\tilde{Z}_{pq}$ the periodic impedance matrix describing the influence of basis functions of surface $q$ on testing functions of surface $p$, $\mathbf{b}_p$ the exciting fields on surface $p$, where indices $1$, $s$ and $2$ stand for the lower interface of the layer, the surface of the inclusions and the upper interface of the layer, respectively. The $\varphi_x$ and $\varphi_y$ dependency of the periodic impedance matrices have been omitted for the sake of compactness. The $\tilde{Z}_{pp}$ blocs of the impedance matrix are composed of two terms, one for each of the two media that the surface $p$ is separating. Similarly, the exciting fields on one surface correspond to the difference between the exciting fields on one side of the surface and the other. Therefore, for $p=1,2$, we can explicitly decompose the total excitation into its two contributions: $\mathbf{b}_p = \mathbf{b}_p^\downarrow-\mathbf{b}_p^\uparrow$, $\mathbf{b}_p^\downarrow$ and $\mathbf{b}_p^\uparrow$ being the value of the exciting fields from above and below the surface, respectively.

The excitations $\mathbf{b}_1^\downarrow$, $\mathbf{b}_2^\uparrow$ and $\mathbf{b}_s$ are due to sources situated inside the layer, while the other ones are due to sources situated outside the layer. The eigenmodes of the structure are solutions of the homogeneous Maxwell's equations and therefore can exist without internal excitation. Since we are only interested in the eigenmodes of the structure, we will not consider sources located inside the layer, thus the internal excitation is set to zero, which produces:
\begin{equation}
\mathbf{b}_1 = -\mathbf{b}_1^\uparrow ~~~~~~ \mathbf{b}_2 = \mathbf{b}_2^\downarrow ~~~~~~ \mathbf{b}_s = \mathbf{0}
\end{equation}

The unknowns describing the equivalent currents on the surface of the inclusions can be analytically removed from the system of equations \cite{6}, leading to

\begin{equation}
\label{eq:Layer_without_object}
\begin{pmatrix}
 \tilde{\mathcal{Z}}_{11} &  -\tilde{\mathcal{Z}}_{12} \\
-\tilde{\mathcal{Z}}_{21} &   \tilde{\mathcal{Z}}_{22}
\end{pmatrix} \cdot
\begin{pmatrix}
\mathbf{x}_1 \\
\mathbf{x}_2
\end{pmatrix}
= -
\begin{pmatrix}
 \mathbf{b}_1 \\
 \mathbf{b}_2
 \end{pmatrix}
\end{equation}

with
\begin{subequations}
\begin{align}
\tilde{\mathcal{Z}}_{11} &= \tilde{Z}_{11} - \tilde{Z}_{1s} \tilde{Z}_{ss}^{-1} \tilde{Z}_{s1} \label{eq:24-02-16-01}\\
\tilde{\mathcal{Z}}_{12} &= \tilde{Z}_{12} - \tilde{Z}_{1s} \tilde{Z}_{ss}^{-1} \tilde{Z}_{s2}\\
\tilde{\mathcal{Z}}_{21} &= \tilde{Z}_{21} - \tilde{Z}_{2s} \tilde{Z}_{ss}^{-1} \tilde{Z}_{s1} \\ 
\label{eq:24-02-16-02}
\tilde{\mathcal{Z}}_{22} &= \tilde{Z}_{22} - \tilde{Z}_{2s} \tilde{Z}_{ss}^{-1} \tilde{Z}_{s2} 
\end{align}
\end{subequations}

\begin{figure}[h!]
\center
\includegraphics[width = 5cm]{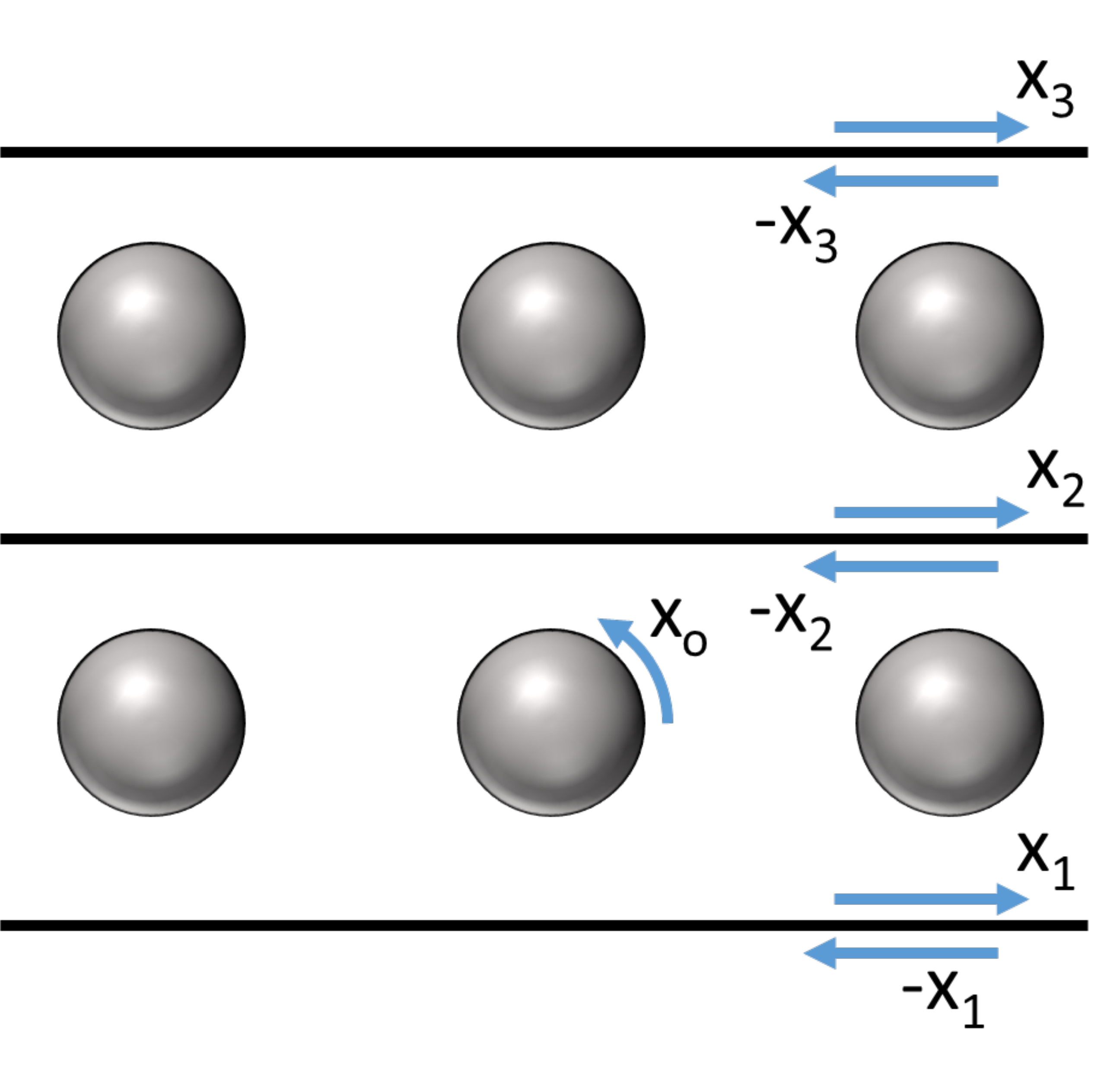}
\caption{Two consecutive layers of the infinite 3D metamaterial. The influence of any source located outside these two layers on interface 2 is screened by the equivalent currents on interfaces 1 and 3.}
\label{fig:2layers}
\end{figure}

We now consider the full 3D metamaterial and search for its eigenmodes, each of them being characterized by its equivalent currents distribution on the lower interface of the layer and a complex wave vector $\mathbf{k}$, such that the fields and currents have the same periodicity as the structure when multiplied by $\exp(j\mathbf{k}\cdot \mathbf{r})$. Note that $\mathbf{k}$ may be complex. Its real part corresponds to a linear phase shift of the mode from one unit cell to the next and the imaginary part corresponds to a change in its amplitude. The transverse wave-vector components $k_x$ and $k_y$ have been set when computing the periodic impedance matrices: $k_x = \varphi_x/d_x$, $k_y = \varphi_y/d_y$. For such a mode, the relation between the equivalent currents on consecutive planes is given by

\begin{equation}
\label{eq:periodic}
\mathbf{x}_{q+1} = g ~\mathbf{x}_q,  ~~~~~~ g = \exp(-j h \mathbf{k} \cdot \hat{z})
\end{equation}

with $h$ the thickness of the layer and $\mathbf{x}_q$ the currents distribution on the lower interface of the $q^{th}$ layer. These eigenmodes correspond to the Bloch modes of the structure. The field distribution inside one layer only depends on the currents on the layer interfaces. Hence imposing the tangential fields continuity across one interface only requires to consider two consecutive layers (cf. Figure \ref{fig:2layers}). The two layers being identical, translational symmetry can be used:

\begin{equation}
\tilde{\mathcal{Z}}_{(p+1)(q+1)}^{\text{L2}} = \tilde{\mathcal{Z}}_{pq}, ~~~~~~ p,q \in \{1,2\}
\end{equation}

with $\tilde{\mathcal{Z}}^{\text{L2}}$ the impedance matrices describing the Layer 2 in the absence of Layer 1. Using (\ref{eq:Layer_without_object}), the continuity of the fields is expressed as

\begin{equation}
\label{eq:21-04_1}
-\tilde{\mathcal{Z}}_{21} \mathbf{x}_1 + \big( \tilde{\mathcal{Z}}_{22} + \tilde{\mathcal{Z}}_{11} - \tilde{Z}_{11} \big) \mathbf{x}_2 - \tilde{\mathcal{Z}}_{12} \mathbf{x}_3 = \mathbf{0}
\end{equation}

The third term in the parenthesis appears due to the fact that the self-interaction of Plane 2 through the homogeneous media above and below the plane is implicitly taken into account twice, since it appears in both $\tilde{\mathcal{Z}}_{11}$ and $\tilde{\mathcal{Z}}_{22}$ (see (\ref{eq:24-02-16-01}) and (\ref{eq:24-02-16-02}), with $\tilde{Z}_{22} = \tilde{Z}_{11}$). The repetition of $\tilde{Z}_{11}$ must therefore be removed.
From (\ref{eq:21-04_1}) and using (\ref{eq:periodic}), the eigenmodes inside the structure satisfy the following equation:

\begin{equation} 
\label{eq:nonlin}
\big( Y_{21} + g^2 Y_{12} \big) \mathbf{x}_2 = g \mathbf{x}_2
\end{equation}

with
\begin{subequations}
\begin{align}
Y_{21} = \big( \tilde{\mathcal{Z}}_{11} + \tilde{\mathcal{Z}}_{22} - \tilde{Z}_{11} \big)^{-1} \tilde{\mathcal{Z}}_{21} \\
Y_{12} = \big( \tilde{\mathcal{Z}}_{11} + \tilde{\mathcal{Z}}_{22} - \tilde{Z}_{11} \big)^{-1} \tilde{\mathcal{Z}}_{12} 
\end{align}
\end{subequations}

Equation (\ref{eq:nonlin}) does not correspond to a basic eigenvalue equation because the matrix $(Y_{21} + g^2 Y_{12} )$, whose eigenvalues are being computed, depends on the eigenvalues themselves, through the $g^2$ factor. This results in a non-linear formulation of the problem, where the terms $g^2 Y_{12}$ and $Y_{21}$ describe the influence of $\mathbf{x}_3$ and $\mathbf{x}_1$ on the interface 2, respectively. 

The eigenvalue problem of Equation (\ref{eq:nonlin}) can be linearized by doubling the number of unknowns of the system of equations. Starting from (\ref{eq:nonlin}), it can be rearranged to obtain

\begin{equation}
Y_{21} \mathbf{x}_2 + g Y_{12} \mathbf{x}_3 = \mathbf{x}_3
\end{equation}

with $\mathbf{x}_2$ and $\mathbf{x}_3$ that are both vectors of unknowns. The system of equations being underdetermined, we can also use the periodicity conditions (\ref{eq:periodic}). We can then write

\begin{equation}
\label{eq:17-05-1}
\begin{pmatrix}
0 & \mathcal{I} \\
-Y_{12}^{-1} Y_{21} & Y_{12}^{-1} 
\end{pmatrix}
\cdot
\begin{pmatrix}
\mathbf{x}_2 \\ \mathbf{x}_3
\end{pmatrix}
= g
\begin{pmatrix}
\mathbf{x}_2 \\ \mathbf{x}_3
\end{pmatrix}
\end{equation}

with $\mathcal{I}$ the identity matrix. However, $Y_{12}^{-1} = \tilde{\mathcal{Z}}_{12}^{-1}  \big( \tilde{\mathcal{Z}}_{11} + \tilde{\mathcal{Z}}_{22} - \tilde{Z}_{11} \big)$ factor appears. Inverting the matrix $\tilde{\mathcal{Z}}_{12}$ physically means that the fields are being back-propagated, which is numerically unstable. This problem also appears with methods that involve the T-matrix. As will be seen in Section \ref{sec:numRes}, it dramatically decreases the accuracy of the results. In the next Section, we propose an alternative formulation that is linear and free of back-propagation and is therefore numerically stable.

This method is based on the fact that, due to energy conservation, in the presence of losses, the $g$ factor has to be smaller than unity. Matrices of Equation (\ref{eq:nonlin}) can be modified to reformulate the problem in the form 

\begin{equation}
\label{eq:21-12-01}
\big( Y_{21}^{(N)} + g^{N} Y_{12}^{(N)} \big) \mathbf{x}_2 = g \mathbf{x}_2
\end{equation}

with $N>2$. If the spectral radius of matrix $Y_{12}^{(N)}$ remains limited for any $N$ bigger than a given threshold, it can be proven that the impact of the second term of the left-hand side of Equation (\ref{eq:21-12-01}) on the final results tends to vanish for $N$ big enough. This assumption will be proven in Section \ref{sec:convergence}.

It is interesting to mention that finding the eigenmodes inside a structure corresponds to a linear eigenvalue problem if the M\"{u}ller formulation \cite{Muller} is used instead of the PMCHWT. Indeed, currents can be determined considering only one layer and two interfaces, such that no $g^2$ term appears in the final problem. However, it is proved in Appendix \ref{sec:muller} that finding eigenmodes using M\"{u}ller formulation also implies back-propagation of the fields and is therefore numerically unstable.  

\section{Linear formulation}
\label{sec:linform}
Solving (\ref{eq:nonlin}) without resorting to back-propagation requires the search for the eigenvalue in the complex plane and may require a high number of eigenvalue decompositions of the matrix $ Y_{21} + g^2 Y_{12}$. Moreover, the procedure must be applied separately for every eigenmode.
However, the impedance matrices obtained from simulations can be manipulated to reformulate the problem and make it linear versus the factor $g$, such that all the eigenmodes of the structure can be determined at once without resorting to back-propagation.

Consider a stack made of two layers. If the structure is excited by sources outside the two layers, the MoM system of equations becomes

\begin{equation}
\label{eq:18}
\begin{pmatrix}
\tilde{\mathcal{Z}}^{(0)}_{11} & -\tilde{\mathcal{Z}}^{(0)}_{12} & 0 \\
-\tilde{\mathcal{Z}}^{(0)}_{21} & \tilde{\mathcal{Z}}^{(0)}_{22} & -\tilde{\mathcal{Z}}^{(0)}_{12} \\
0 & -\tilde{\mathcal{Z}}^{(0)}_{21} & \tilde{\mathcal{Z}}^{(0)}_{33}
\end{pmatrix}
\begin{pmatrix}
\mathbf{x}_1 \\
\mathbf{x}_2 \\
\mathbf{x}_3
\end{pmatrix}
=
-\begin{pmatrix}
\mathbf{b}_1 \\
\mathbf{0} \\
\mathbf{b}_3
\end{pmatrix}
\end{equation} 

with $\tilde{\mathcal{Z}}^{(0)}_{pq} = \tilde{\mathcal{Z}}_{pq}$ for $p\neq q$ and

\begin{subequations}
\begin{align}
\tilde{\mathcal{Z}}^{(0)}_{11} &= \tilde{\mathcal{Z}}_{11} \\
\tilde{\mathcal{Z}}^{(0)}_{22} &= \tilde{\mathcal{Z}}_{11} + \tilde{\mathcal{Z}}_{22} - \tilde{Z}_{11} \\
\tilde{\mathcal{Z}}^{(0)}_{33} &= \tilde{\mathcal{Z}}_{22} 
\end{align}
\end{subequations}

From there, it is possible to eliminate the unknowns associated to equivalent surface currents on the central interface. We then obtain a system of equations that is directly relating currents on the two outer interfaces to the excitation induced on them. We can then stack two of these double layers to obtain the new system of equations describing 4 layers. The resulting system of equations reads:

\begin{equation}
\begin{pmatrix}
\tilde{\mathcal{Z}}^{(1)}_{11} & -\tilde{\mathcal{Z}}^{(1)}_{12} & 0 \\
-\tilde{\mathcal{Z}}^{(1)}_{21} & \tilde{\mathcal{Z}}^{(1)}_{22} & -\tilde{\mathcal{Z}}^{(1)}_{12} \\
0 & -\tilde{\mathcal{Z}}^{(1)}_{21} & \tilde{\mathcal{Z}}^{(1)}_{33}
\end{pmatrix}
\begin{pmatrix}
\mathbf{x}_1 \\
\mathbf{x}_3 \\
\mathbf{x}_5
\end{pmatrix}
=
-\begin{pmatrix}
\mathbf{b}_1 \\
\mathbf{0} \\
\mathbf{b}_5
\end{pmatrix}
\end{equation} 

with 
\begin{subequations}
\label{eq:V2:1}
\begin{align}
\tilde{\mathcal{Z}}^{(i+1)}_{11} &= \tilde{\mathcal{Z}}^{(i)}_{11} - \tilde{\mathcal{Z}}^{(i)}_{12}\Big(\tilde{\mathcal{Z}}^{(i)}_{22}\Big)^{-1} \tilde{\mathcal{Z}}^{(i)}_{21} \\
\tilde{\mathcal{Z}}^{(i+1)}_{33} &= \tilde{\mathcal{Z}}^{(i)}_{33} - \tilde{\mathcal{Z}}^{(i)}_{21}\Big(\tilde{\mathcal{Z}}^{(i)}_{22}\Big)^{-1} \tilde{\mathcal{Z}}^{(i)}_{12} \\
\tilde{\mathcal{Z}}^{(i+1)}_{22} &= \tilde{\mathcal{Z}}^{(i+1)}_{11} + \tilde{\mathcal{Z}}^{(i+1)}_{33} - \tilde{Z}_{11} \\
\tilde{\mathcal{Z}}^{(i+1)}_{12} &=  \tilde{\mathcal{Z}}^{(i)}_{12}\Big(\tilde{\mathcal{Z}}^{(i)}_{22}\Big)^{-1} \tilde{\mathcal{Z}}^{(i)}_{12} \\
\tilde{\mathcal{Z}}^{(i+1)}_{21} &=  \tilde{\mathcal{Z}}^{(i)}_{21}\Big(\tilde{\mathcal{Z}}^{(i)}_{22}\Big)^{-1} \tilde{\mathcal{Z}}^{(i)}_{21} 
\end{align}
\end{subequations}

Equations (\ref{eq:V2:1}) describe a relation of recurrence that relates the matrices that appear when $2^{i+1}$ layers are treated to those used for $2^i$ layers. Iterating $n$ times, one can rapidly obtain the system of equations describing the link between currents on interfaces separated by $2^n$ layers and the external excitation. If losses are introduced in the structure, when interfaces will be far apart, their mutual coupling will tend to vanish. The number of layers growing exponentially, the losses introduced can be made arbitrarily small in order to influence as little as possible the final results.
%
%From a physical point of view, the solution we are seeking after is a solution that takes into account the scattering of the fields generated by $\mathbf{x}_1$ by the infinite array of spheres located above it. The scattering by the spheres located within the $2^{i}$ first layers is explicitly taken into account via the $\tilde{\mathcal{Z}}^{(i)}_{11}$ matrix, while the impact of the rest of them is taken implicitly into account by the equivalent currents on the top of the slab. Upon each iteration, the impact of scattering by the spheres of the $2^{i-1}$ next layers, which were implicitly taken into account by the $\mathbf{x}_3$ screening currents, is removed from these currents and added explicitly into the $\tilde{\mathcal{Z}}^{(i)}_{11}$ matrix. Considering that the fields generated by the distant spheres onto the lower plane vanish due to losses when the distance between the spheres and the plane is large enough, the solutions of the simplified eigenvalue equation where the $\mathbf{x}_3$ term has been neglected converges towards the solutions of the exact eigenvalue problem. This will be demonstrated in the next Section.

From a physical point of view, the solution we are seeking after is a solution that takes into account the fields generated by $\mathbf{x}_1$ and scattered by the infinite array of spheres located above it. The scattering by the spheres located within the $2^{i}$ first layers is explicitly taken into account via the $\tilde{\mathcal{Z}}^{(i)}_{11}$ matrix, while the impact of the rest of them is taken implicitly into account by the equivalent currents on the top of the slab. Upon each iteration, the impact of scattering by the spheres of the $2^{i-1}$ next layers, which were implicitly taken into account by the screening currents of the upper interface, is removed from these currents and added explicitly into the $\tilde{\mathcal{Z}}^{(i)}_{11}$ matrix. Considering that the fields generated by the distant spheres onto the lower plane vanish due to losses when the distance between the spheres and the plane is large enough, the solutions of the simplified eigenvalue equation where the impact of the currents on the upper plane (i.e. of the distant spheres) has been neglected converges towards the solutions of the exact eigenvalue problem. This will be demonstrated in the next Section.

Considering that after $n$ iterations, the distance between the two outer interfaces is large enough, so that any field scattered by spheres that have not been explicitly taken into account is negligible on the lower interface, it leads to 

\begin{subequations}
\begin{align}
\tilde{\mathcal{Z}}^{(n)}_{12} \simeq \tilde{\mathcal{Z}}^{(n)}_{21} &\simeq 0\\
\mathbf{x}_1 &\simeq -\Big(\tilde{\mathcal{Z}}^{(n)}_{11}\Big)^{-1} \mathbf{b}_1 
\end{align}
\end{subequations}

\begin{figure}[h!]
\center
\includegraphics[width = 5.3cm]{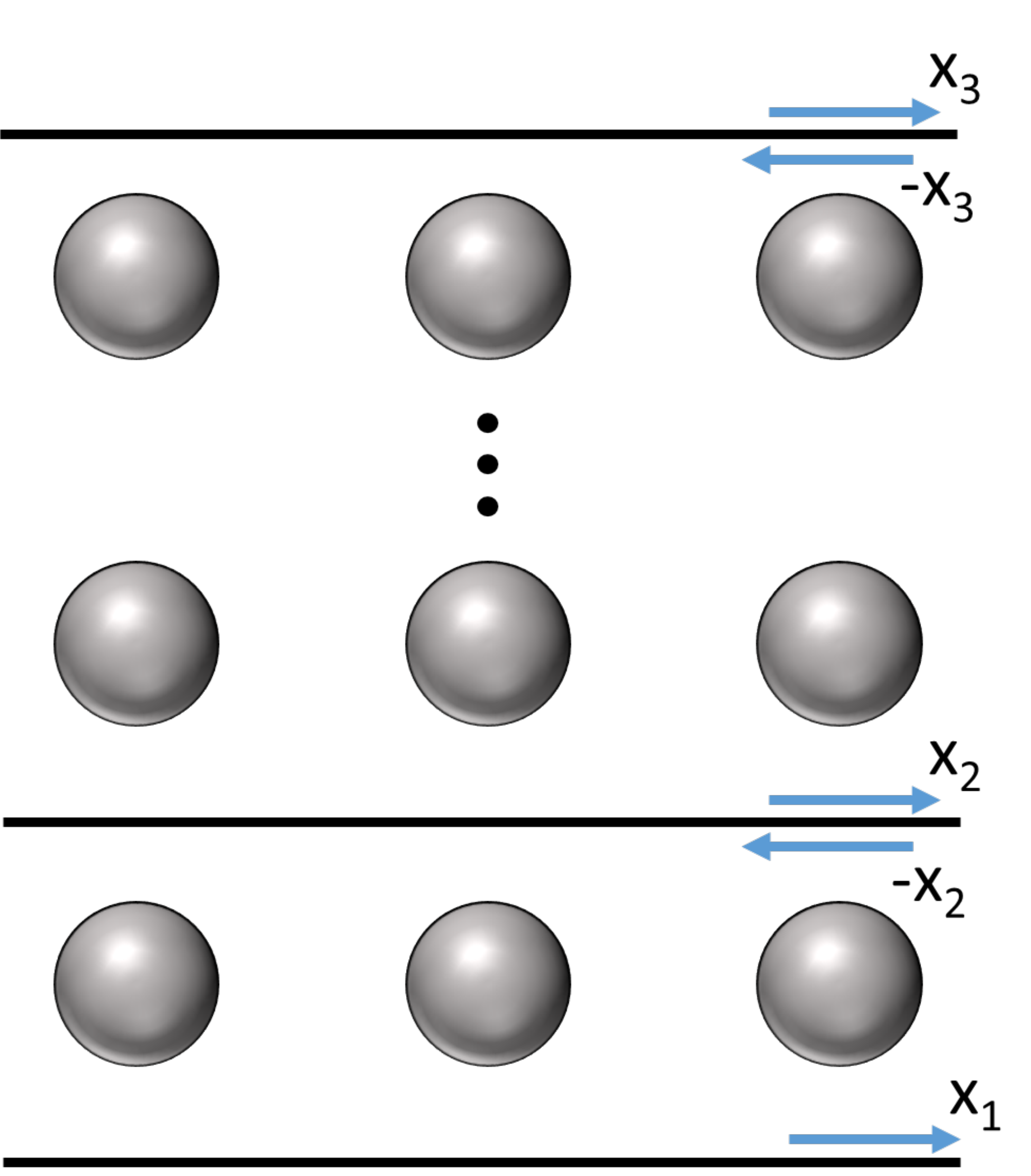}
\caption{Structure considered when computing the eigenmodes of the structure. The number of layers separating interfaces 2 and 3 is equal to $2^n$, $n$ being the number of iterations.}
\label{fig:final_disp}
\end{figure}

The non-linearity of Equation (\ref{eq:nonlin}) arose from the influence of the currents on the upper plane. It can be seen that the problem can now be linearized. Indeed, adding one more layer below the $2^n$ layers that have been considered previously (cf. Figure \ref{fig:final_disp}) and imposing the continuity of the tangential fields across interface 2, one obtains:
\begin{equation}
\label{eq:28-05-03}
\mathbf{x}_2 = -\big(\tilde{\mathcal{Z}}_{11}^{(n)}\big)^{-1} \mathbf{b}_2
\end{equation}
with
\begin{subequations}
\begin{align}
\mathbf{b}_2 =& -\tilde{\mathcal{Z}}_{21} \mathbf{x}_1 - \tilde{\mathcal{Z}}_{12}^{(n)} \mathbf{x}_3 \label{eq:28-05-02}\\
\simeq& -\tilde{\mathcal{Z}}_{21} \mathbf{x}_1
\end{align}
\end{subequations}

Using (\ref{eq:periodic}), it results in:
\begin{equation}
\label{eq:06-03-5}
\Big( \tilde{\mathcal{Z}}_{11}^{(n)} \Big)^{-1} \tilde{\mathcal{Z}}_{21} \mathbf{x}_1 \simeq g \mathbf{x}_1
\end{equation}

All the eigenmodes and eigenvectors can now be found directly applying an eigenvalue decomposition of the matrix in the left-hand side. 

The advantage of this method is that the method is numerically stable and each iteration only requires the inversion of one block of the impedance matrix and few matrix-matrix multiplications. Moreover, the number of layers growing exponentially, the convergence is very rapid. Generally, the time needed to retrieve all the eigenmodes starting from (\ref{eq:8-05-1}) is negligible compared to that needed to compute the impedance matrices appearing in (\ref{eq:8-05-1}).

\section{Convergence of the method}
\label{sec:convergence}

The method we presented is based on the hypothesis that the fields excited on one side of a slab of lossy metamaterial will tend to decrease when traveling to the other end of the slab, even considering multiple reflections at the interfaces, provided that the number of layers is large enough. The advantage of the proposed method is that the number of layers grows exponentially with the number of iterations, such that such large numbers can be rapidly obtained. 

To estimate the number of iterations required, it is necessary to evaluate the convergence of the method and the influence of losses.
In this section, we show that the convergence of the proposed method with respect to the number of iterations and for a given level of losses is doubly exponential, leading to the rapid convergence of the results. Moreover, the number of iterations required for a given level of accuracy increases only linearly for exponentially decreasing losses.

Consider a mode $\alpha$ traveling through a slab of metamaterial made of $N$ layers. This mode is characterized by the fields distribution within one layer $(\mathbf{E}_\alpha(\mathbf{r}), \mathbf{H}_\alpha(\mathbf{r}))$ and the transmission coefficient $g_\alpha$. The fields distribution is normalized, such that 

\begin{equation}
\label{eq:24-02-16-03}
\iint_{S} (\|\mathbf{E}_\alpha \|_2^2 + \|\eta\mathbf{H}_\alpha\|_2^2) dS = 1
\end{equation}

with $S$ the surface of the lower interface of the layer limited to one unit cell, $\eta$ the impedance of the medium just above that surface and $\|\mathbf{A}\|_2$ the L2 norm of vector $\mathbf{A}$. Note that this normalization has been chosen arbitrarily, but any other normalization might have been chosen instead. In the following, the direction of propagation of a mode is defined so that it corresponds to the direction of the energy flow associated to that mode at each interface.

Using energy conservation, it is possible to prove that the magnitude of the transmission coefficient $|g_\alpha|$ has to be smaller than unity provided that the metamaterial is lossy: the power due to one mode crossing two consecutive interfaces is constant within a $|g_\alpha^2|$ factor. The difference in the power entering one layer and the power exiting that same layer corresponds to the dissipations, which are strictly positive for a lossy medium. It provides:
\begin{equation}
\label{eq:01-06-06}
|g_\alpha| < 1
\end{equation}
Moreover if losses are introduced using a non-zero conductivity $\sigma$ that is small enough, the transmission coefficient can be expressed as a function of $\sigma$ using first-order Taylor series:
\begin{equation}
\label{eq:28-05-08}
|g_\alpha| \simeq 1-\rho_\alpha \sigma
\end{equation}
with $\rho_\alpha$ a constant that depends on the geometry of the structure and the fields distribution of the mode when there are no losses. A more formal derivation of these results and a physical insight on $\rho_\alpha$ can be found in Appendix \ref{app:B}.

Now, we consider the error in the fields due to the finiteness of the metamaterial slab represented in Fig. \ref{fig:final_disp}.
Substituting Equation (\ref{eq:28-05-02}) into (\ref{eq:28-05-03}), one can find that the exact currents $\mathbf{x}_2$ are given by

\begin{equation}
\label{eq:28-05-05}
\mathbf{x}_2 = \big(\tilde{\mathcal{Z}}_{11}^{(n)}\big)^{-1} \Big( \tilde{\mathcal{Z}}_{21} \mathbf{x}_1 + \mathcal{\tilde{Z}}_{12}^{(n)} \mathbf{x}_3 \Big)
\end{equation}

The solution, that is found combining (\ref{eq:periodic}) and (\ref{eq:28-05-05}), corresponds to one of the eigenmodes of the structure. Without any loss of generality, we consider that this mode is the mode $\alpha$ of the structure.
This mode is excited by currents $\mathbf{x}_1$ at the lowest interface of the structure and its amplitude is 1 on interface 2 (see Fig. \ref{fig:final_disp}). The power transported by that mode crossing interface $p$, between layers $p-1$ and $p$, is given by

\begin{equation}
P_\alpha^{p} = \Re\Big\{ \iint_{S_p} \dfrac{\mathbf{E}_\alpha(\mathbf{r}) \times \mathbf{H}_\alpha^*(\mathbf{r})}{2} \cdot \hat{n} ~ dS \Big\}
\end{equation}

with $\Re\{A\}$ the real part of $A$, $\hat{n}$ the unit vector normal to interface $p$, toward layer $p$, and $S_p$ the area of that interface limited to one unit-cell. Using Equation (\ref{eq:periodic}) the total power crossing interface $p$ can be related to the power crossing Plane 0:
\begin{equation}
\label{eq:01-06-01}
P_\alpha^{p} = |g_\alpha|^{2p} \Re\Big\{  \iint_{S_0} \dfrac{\mathbf{E}_\alpha(\mathbf{r}) \times \mathbf{H}_\alpha^*(\mathbf{r})}{2} \cdot \hat{n} ~ dS \Big\}
\end{equation}

Now, we consider what happens in Plane 3 where currents $\mathbf{x}_3$ are flowing. The amplitude of mode $\alpha$ after propagating through $N+1$ layers is $g_\alpha^{N}$. If we neglect the term due to currents $\mathbf{x}_3$ in (\ref{eq:28-05-05}), this mode will be scattered at the interface between the metamaterial and the air. It means that the energy transported by this mode will be partially transmitted into air and partially reflected back into the metamaterial. The part of the energy that is reflected at the interface will spread over many different propagating and evanescent modes.
We consider here the worst case, where all the power transported by mode $\alpha$ would be reflected back into the metamaterial and redirected into a single mode that we will call $\beta$. As a worst-case, 
that mode is chosen as the mode inducing the biggest fields on Plane 2 for a given amplitude of mode $\alpha$ in Plane 3 (cf. Fig. \ref{fig:final_disp}). 
%this mode is chosen as the most propagative one, i.e. the mode whose transmission coefficient $g_\beta$ is the closest to unity among all the possible modes.
If we define $C_{\beta, \alpha}$ as the scattering coefficient of mode $\alpha$ into mode $\beta$,  using energy balance in Plane 3, we can find that:

\begin{equation}
\label{eq:06-01-01}
|C_{\beta, \alpha}|^2 = \dfrac{P_\alpha}{P_\beta} = -
\dfrac{\Re\{\iint_{S_0} {\mathbf{E}_\alpha(\mathbf{r}) \times \mathbf{H}_\alpha^*(\mathbf{r})} \cdot \hat{n} ~ dS\}}
{\Re\{\iint_{S_0} {\mathbf{E}_\beta(\mathbf{r}) \times \mathbf{H}_\beta^*(\mathbf{r})} \cdot \hat{n} ~ dS\}}
\end{equation}

Note that using this equation, only the amplitude of $C_{\beta, \alpha}$ can be determined, not its phase. The mode $\beta$ generated in Plane 3 will propagate back to Plane 1, where its amplitude will be reduced by a factor $g_\beta^{N+1}$. There, the same worst-case reasoning can be applied, considering that all the energy transported by mode $\beta$ is reflected back into the metamaterial and entirely contained into a single mode $\gamma$. That mode is chosen as the worst possible case, i.e. the mode inducing the biggest fields on Plane 2 for a given amount of power transported through Plane 1. This mode $\gamma$ will propagate until it reaches Plane 3. There, again, we consider that all the energy transported by mode $\gamma$ is reflected back into the metamaterial and is entirely contained in the worst possible mode, which we previously defined as $\beta$. Iterating the same procedure for an infinite number of reflections at the interfaces, one can find the total amplitudes $K_\beta$ and $K_\gamma$ of modes $\beta$ and $\gamma$, respectively, on Plane 2, for a unit amplitude of mode $\alpha$ on Plane 2:

\begin{subequations}
\label{eq:24-12-01}
\begin{align}
K_\beta &= g_\alpha^{N}  \sum_{q=0}^{\infty} C_{\beta, \alpha} g_\beta^{N} \Big( C_{\gamma, \beta} C_{\beta, \gamma} g_\beta^{N+1} g_\gamma^{N+1} \Big)^q \\
K_\gamma &= g_\alpha^{N}  \sum_{q=0}^{\infty} C_{\beta, \alpha} C_{\gamma, \beta} g_\beta^{N+1} g_\gamma \Big( C_{\gamma, \beta} C_{\beta, \gamma} g_\beta^{N+1} g_\gamma^{N+1} \Big)^q 
\end{align}
\end{subequations}

Using (\ref{eq:01-06-06}) and (\ref{eq:06-01-01}), one can notice that Equation (\ref{eq:24-12-01}) corresponds to a geometrical series whose terms are smaller than unity. It yields

\begin{subequations}
\label{eq:06-01-02}
\begin{align}
K_\beta &= g_\alpha^{N}  \dfrac{C_{\beta, \alpha} g_\beta^{N}}{1- \exp(i\varphi) g_\beta^{N+1} g_\gamma^{N+1} }\\
K_\gamma &= g_\alpha^{N}  \dfrac{C_{\beta, \alpha} C_{\gamma, \beta} g_\beta^{N+1} g_\gamma}{1- \exp(i\varphi) g_\beta^{N+1} g_\gamma^{N+1} }
\end{align}
\end{subequations}

with $\varphi$ the phase of the $C_{\gamma, \beta} C_{\beta, \gamma}$ term, whose magnitude is 1 according to (\ref{eq:06-01-01}). Noting that $C_{i,j}$ has to be of finite amplitude and that $|g_i^N|$ can be made arbitrarily small by increasing the number of layers, we see that if the number of layers grows, (\ref{eq:06-01-02}) converges toward $K_\beta = K_\gamma = 0$. Therefore, for a growing number of layers and considering that the medium is lossy, we can see that multiple reflections at the interfaces become negligible and the proposed method converges toward the right solution. This proof can be generalized to the case of multiple propagative modes. The scattering coefficients $C_{i,j}$ then become matrices of finite spectral radius. Using the same reasoning, we can demonstrate that it does not change the convergence of the solution, i.e. $K_\alpha = 1, ~K_\delta = 0, ~\forall \delta \neq \alpha$.

An interesting digression here concerns phenomena such as extraordinary transmission through a layered medium that strongly depends on the number of layers considered \cite{christmas1}. Such phenomena are due to multiple reflections on both interfaces of the metamaterial slab and the air. Mathematically, this is due to the denominator of (\ref{eq:06-01-02}) that may vary significantly with the phase of the $\exp(i \varphi) g_\beta^{N+1} g_\gamma^{N+1}$ term. However, in the presence of losses, these variations together with the multiple reflections become negligible. Therefore, phenomena like the one described in \cite{christmas1} disappear and the eigenmodes of the structure can be retrieved, provided that the number of layers considered is large enough.

%{\color{magenta}
%An interesting digression here concerns phenomena such as extraordinary transmission through a layered medium that strongly depends on the number of layers considered \cite{christmas1}. Such phenomena are due to multiple reflections on both interfaces of the metamaterial slab and the air. While the eigenmodes of the structure seemingly depend on the number of layers, it is actually the linear combination of these modes that is solution of the problem that strongly varies with the number of layers. However, in the presence of losses, this coupling between the two interfaces tends to vanish and the eigenmodes of the structure can be retrieved whatever is the number of layers considered provided that it is large enough.
%}

Now that we have proven the convergence of the method, we will give an upper bound for the error with respect to the number of iterations. We define the error on the fields for a number $N$ of layers, i.e. the difference between the exact solution and the approximate one which contains some $\beta$ and $\gamma$ modes, as

\begin{equation}
\label{eq:24-02-16-04}
e =  \dfrac{\iint_{S_1} \big\|\mathbf{f}_\alpha - (\mathbf{f}_\alpha + K_\beta \mathbf{f}_\beta + K_\gamma \mathbf{f}_\gamma ) \big\|_2 dS}{\iint_{S_1} \|\mathbf{f}_\alpha\|_2  dS}
\end{equation}
with $\mathbf{f}_i = (\mathbf{E}_i, \eta \mathbf{H}_i)$ the fields on the interface between two layers due to mode $i$ of unit amplitude.
In the following, we will consider that the solution has converged if the error $e$ defined by (\ref{eq:24-02-16-04}) is smaller than a predefined threshold $e_0 \ll 1$.
Going back to the definition of the modes (\ref{eq:24-02-16-03}), it can be seen that $\iint_{S_1} \|\mathbf{f}_i \|_2 dS = 1$. We then obtain

\begin{equation}
\label{eq:14-03-16-01}
e \leq |K_\beta| + |K_\gamma|
\end{equation}

As explained previously, the sums in Equation (\ref{eq:24-12-01}) are converging to zero for an increasing number of layers $N$. The sums correspond to geometrical series, such that it is absolutely convergent \cite{convergence}.
% If we postulate that $N$ is big enough for the {\color{red} magnitude of the sums} to be smaller than $1/2$. 
 We can hence consider that there exists an integer $N_0$ such that the magnitude of the sums in (\ref{eq:24-12-01}) is smaller than  $1/2$ for any number of layers larger than $N_0$. From there, two possibilities:
\begin{itemize}
\item The mode $\alpha$ is highly evanescent. In that case, the magnitude of the term $g_\alpha^N$ in Equations (\ref{eq:24-12-01}) will decrease much faster than the magnitude of the sums. The convergence may therefore be reached for a number of layers $N$ that is smaller than $N_0$.
\item The mode $\alpha$ is propagative or slightly evanescent. In that case, the convergence of the solution will require a number $N$ of layers that is much larger than $N_0$. 
\end{itemize}
In the first case, imposing $N=N_0$ is sufficient to ensure convergence. Indeed, the evanescent aspect of the mode leads to natural convergence.
In the second case, which treats the dominant modes |in which we are primarily interested| a number of layers $N>N_0$ is required to reach convergence.  Substituting (\ref{eq:24-12-01}) into (\ref{eq:14-03-16-01}) and considering that the magnitude of the sums of (\ref{eq:24-12-01}) is smaller than $1/2$ gives:
%In order to estimate the rate of convergence of the method, we will consider that the number of layers $N$ required to reach convergence is larger than $N_0$. This assumption holds true when computing the propagating or slightly evanescent modes of the structure, in which one is generally interested. However, in the case of highly evanescent modes, fewer layers will be needed to reach convergence and the assumption $N>N_0$ may not be satisfied. Therefore, for these modes, the actual convergence rate of the method may be overestimated by the following formulas, while being still faster than the convergence rate of propagative modes. The latter can therefore be used as a lower limit.
%
%However, the convergence rate for these modes will remain faster than the convergence rate of the propagative ones. The latter may therefore be viewed as a lower limit for the convergence rate of the method when computing every eigenmode of the structure.
%the convergence may be reached before this assumption becomes true and therefore the convergence rate may be slightly overestimated in the following. However, it is not a problem as the convergence of these evanescent modes will remain much faster than the convergence of the propagative ones, which will be the ones limiting the convergence speed. 
%
%Making the hypothesis that $N>N_0$, (\ref{eq:14-03-16-01}) becomes}
 
 \begin{equation}
e < |g_\alpha|^N
\end{equation}

with $N=2^n$ the number of layers inside the slab. Imposing that the solution is converged and using (\ref{eq:28-05-08}), we obtain
\begin{equation}
(1-\rho_\alpha \sigma)^{2^n} < e_0
\end{equation}
which can be reformulated as
\begin{equation}
-2^n  \ln(1-\rho_\alpha \sigma) > -\ln(e_0)
\end{equation}
The conductivity being small, the logarithm of the left-hand side can be approximated using first-order Taylor series. Taking the logarithm of the resulting expression, we obtain
\begin{equation}
n + \log_2(\rho_\alpha \sigma) > \log_2 \big(-\ln(e_0) \big)
\end{equation}
After few manipulations, we can find that the error remains below the threshold $e_0$ for any number of iterations $n$ such that:
\begin{equation}
\label{eq:convergence}
n > \log_2\big(\ln(e_0^{-1})\big) - \log_2 \big( \rho_\alpha \sigma \big)
\end{equation}

As we can see, the convergence is exponential with respect to the losses introduced in the material and doubly exponential with respect to the error threshold. 

It is interesting to notice that the shape and constitutive parameters of the inclusions are only impacting the convergence through the factor $\rho_\alpha$ (see (\ref{eq:28-05-08}) and Appendix \ref{app:B}). Indeed, the solutions to Maxwell's equations will change depending on the geometry and constitutive parameters of the inclusions. So will do the dissipations associated to these solutions. 

So far, losses have been introduced via a small conductivity $\sigma$. Another possibility to introduce losses is to numerically decrease the amplitude of the transmission coefficient $g$ by an arbitrary factor $(1-\gamma)$. In that case, the losses, which have no physical origin, are independent of the geometry and constitutive parameters of the inclusions. The number of iterations required to reach a given level of accuracy becomes

\begin{equation}
n > \log_2\big(\ln(e_0^{-1})\big) - \log_2 \big( \gamma \big)
\end{equation}

which is independent of the structure considered. 

Note that the accuracy of the method can be tested a posteriori by substituting the obtained results into the non-linear formulation (\ref{eq:nonlin}).

\section{Comparison with T-matrix based methods}
\label{sec:Smat}
The proposed method has been presented using the MoM formalism as a support. However, the only information that is needed is the way in which fields propagate through one layer. In this section, we extend the method to the more versatile Scattering-matrix (S-matrix) formalism \cite{8}.

The S-matrix of a structure is the matrix that relates the fields incident on the structure to those scattered by the structure. That matrix is generally well conditioned \cite{8}. The Transfer-matrix (T-matrix) of the layer is the matrix that relates the total fields (incident and scattered) on one side of the layer to the total fields on the other side of the layer. Because it has to handle back-propagation, the T-matrix is numerically ill-conditioned \cite{8}. Such as the MoM impedance matrix, S- and T-matrices are different ways to represent the response of a structure to incident fields.

Methods that use the information of S- and T-matrices to determine the eigenmodes of a structure have already been demonstrated in the literature \cite{4,5,limitations}. The response of one layer of the structure studied to incident electromagnetic fields is computed through the computation of the S-matrix associated to that layer. To determine the eigenmodes of the structure from the information stored in the S-matrix, the latter is used to compute the T-matrix of a layer. All the eigenmodes and their corresponding eigenvalues can be computed at once using an eigenvalue decomposition of that T-matrix. However, as shown in \cite{limitations}, the computation of the T-matrix is numerically ill-conditioned and therefore prone to numerical noise. The maximum precision that can be achieved is therefore limited.

Although the method is still relatively robust using plane-wave decomposition of the fields \cite{4}, we found it to be unstable when the decomposition of the fields were done using equivalent currents. We encountered a strong dependence of the results on the mesh used to express the currents distribution and to test the fields using the M\"{u}ller formulation of the MoM \cite{7}. Results obtained for the linear problem with a double number of unknowns (cf. Section \ref{sec:non-lin}) were found to be more stable, but still prone to numerical error propagation, as will be shown in Section \ref{sec:numRes}.

As shown for MoM impedance matrix formalism, this limitation can be circumvented by linearizing the eigenvalue problem associated to the S-matrix without using the equivalent T-matrix.

%%%%%%%%%%%%%%%%%%%%%%%%%%%%%%%%%%%%%
%%%%%%%%%%%%%%%%%%%%%%%%%%%%%%%%%%%%%
%%%%%%%%%%%%%%%%%%%%%%%%%%%%%%%%%%%%%
%%%%%%%%%%%%%%%%%%%%%%%%%%%%%%%%%%%%%
%%%%%%%%%%%%%%%%%%%%%%%%%%%%%%%%%%%%%
%%%%%%%%%%%%%%%%%%%%%%%%%%%%%%%%%%%%%
%%%%%%%%%%%%%%%%%%%%%%%%%%%%%%%%%%%%%
%%%%%%%%%%%%%%%%%%%%%%%%%%%%%%%%%%%%%
%%%%%%%%%%%%%%%%%%%%%%%%%%%%%%%%%%%%%
%%%%%%%%%%%%%%%%%%%%%%%%%%%%%%%%%%%%%
%%%%%%%%%%%%%%%%%%%%%%%%%%%%%%%%%%%%%

\begin{figure}[h!]
\center
\includegraphics[width = 4 cm]{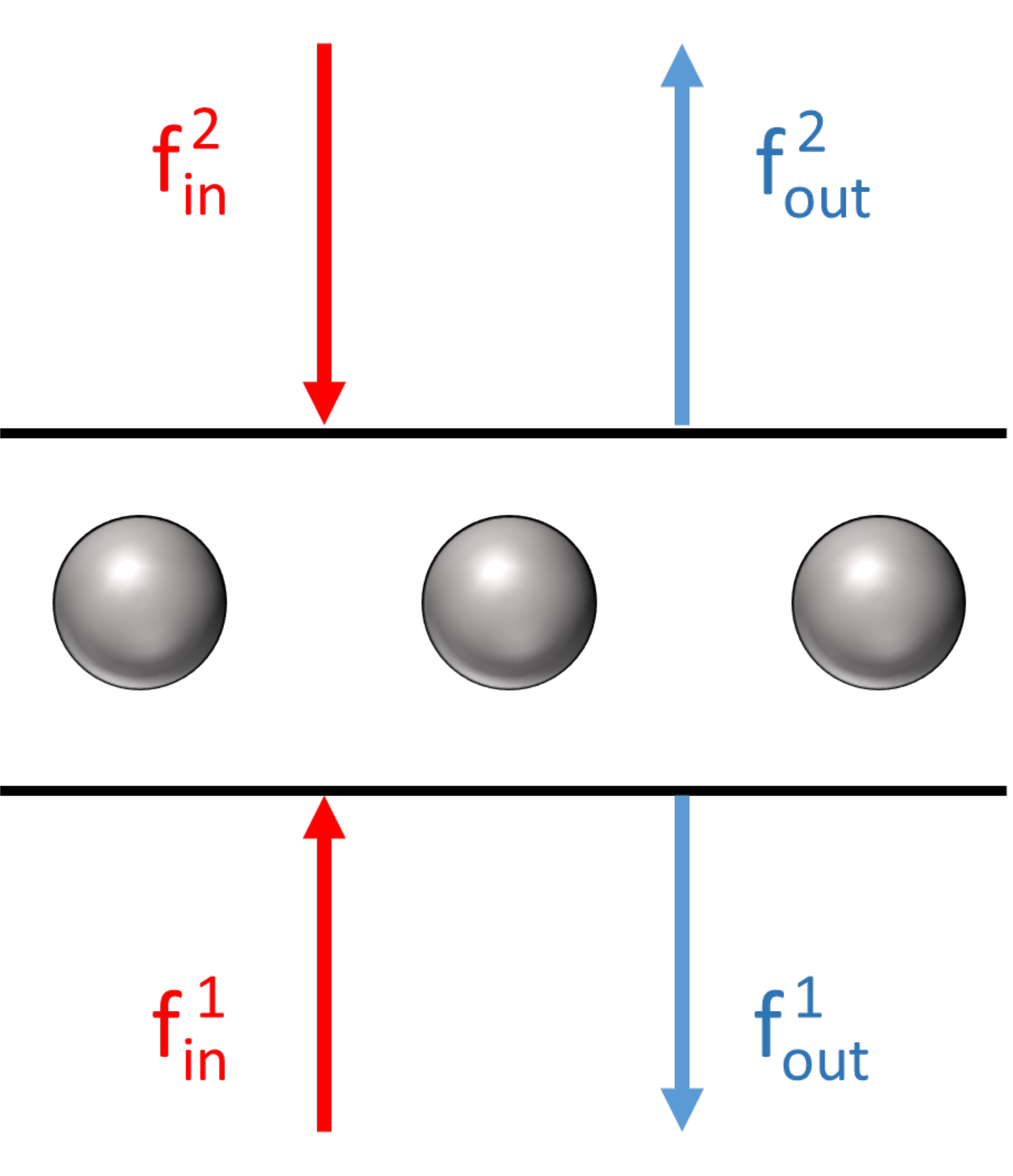}
\caption{Scheme of the S-matrix formalism used. The relation between the incoming fields that excite the layer (in red) and the fields scattered by the layer (in blue) is characterized by the scattering matrix $S^{(0)}$. }
\label{fig:Tmatrix}
\end{figure}

We first consider one layer of the metamaterial. The field scattered by the layer can be related to the fields incident on that layer using the S-matrix $S^{(0)}$ :

\begin{equation}
\label{eq:Smat}
\begin{pmatrix}
S_{11}^{(0)} & S_{12}^{(0)} \\
S_{21}^{(0)} & S_{22}^{(0)} 
\end{pmatrix}
\begin{pmatrix}
\mathbf{f}_{in}^1 \\ \mathbf{f}_{in}^2
\end{pmatrix}
=
\begin{pmatrix}
\mathbf{f}_{out}^1 \\ \mathbf{f}_{out}^2
\end{pmatrix}
\end{equation}

with $\mathbf{f}_{in}^i$ and $\mathbf{f}_{out}^{i}$ the fields incident on and radiated from the layer, respectively, as illustrated in Figure \ref{fig:Tmatrix}. 
The eigenmodes of the metamaterial can then be found using

\begin{subequations}
\label{eq:01-12-01}
\begin{align}
\mathbf{f}_{out}^{1} &= g^{-1} \mathbf{f}_{in}^{2} \\
\mathbf{f}_{out}^{2} &= g \mathbf{f}_{in}^1
\end{align}
\end{subequations}

Substituting these conditions into Equation (\ref{eq:Smat}), one can obtain the non-linear formulation of the problem equivalent to (\ref{eq:nonlin}). A linear formulation can be found using an iterative procedure similar to that described in Section \ref{sec:linform}.

We now add a second layer on top of the first one. The total scattering matrix can be computed for the structure made of two layers:

\begin{equation}
\begin{pmatrix}
S_{11}^{(1)} & S_{12}^{(1)} \\
S_{21}^{(1)} & S_{22}^{(1)} 
\end{pmatrix}
\begin{pmatrix}
\mathbf{f}_{in}^1 \\ \mathbf{f}_{in}^2
\end{pmatrix}
=
\begin{pmatrix}
\mathbf{f}_{out}^1 \\ \mathbf{f}_{out}^2
\end{pmatrix}
\end{equation}

with 

\begin{subequations}
\begin{align}
S_{11}^{(i+1)} &= S_{11}^{(i)} + S_{12}^{(i)} A_{11}^{(i)} S_{21}^{(i)} \\
S_{12}^{(i+1)} &= S_{12}^{(i)} A_{12}^{(i)} S_{12}^{(i)} \\
S_{21}^{(i+1)} &= S_{21}^{(i)} A_{21}^{(i)} S_{21}^{(i)} \\
S_{22}^{(i+1)} &= S_{22}^{(i)} + S_{21}^{(i)} A_{22}^{(i)} S_{12}^{(i)} 
\end{align}
\end{subequations}

and 

\begin{equation}
\begin{pmatrix}
A_{11}^{(i)} & A_{12}^{(i)} \\
A_{21}^{(i)} & A_{22}^{(i)}
\end{pmatrix}
= 
\begin{pmatrix}
-S_{22}^{(i)} & \mathcal{I} \\
\mathcal{I} & -S_{11}^{(i)}
\end{pmatrix}^{-1}
\end{equation}

$\mathcal{I}$ standing for the identity matrix. Note that $\mathbf{f}_{in}^i$ ($\mathbf{f}_{out}^i$) still represents the fields incident on (scattered by) the device, which now consists of a stack of two layers. Iterating the procedure $n$ times, the obtained scattering matrix describes the scattering of incident fields by a metamaterial made of $2^n$ consecutive layers. If the number of layers considered is large enough and some losses are added to the structure, the spectral radius of matrices $S_{12}^{(n)}$ and $S_{21}^{(n)}$ will tend to vanish.
It is worth noting that the operation of adding layers using this method is numerically stable, as demonstrated in \cite{8}.

Once enough layers have been added, we can consider the structure illustrated in Figure \ref{fig:final_disp}. As $\mathbf{f}_{in}^{2}$ has now a negligible impact on resulting $\mathbf{f}_{out}^{1}$, all the quantities are now only depending on $\mathbf{f}_{in}^1$. Simplifying the formula accordingly, one can find the eigenmodes by solving

\begin{equation}
\label{eq:05_03_4}
\Big( \mathcal{I} - S_{22}^{(0)} S_{11}^{(n)} \Big)^{-1} S_{21}^{(0)} \mathbf{f}_{in}^1 = g \mathbf{f}_{in}^1
\end{equation}

Once solved, the total fields distribution associated to the mode can be found knowing that:

\begin{equation}
\mathbf{f}_{out}^1 = S_{11}^{(n)} \mathbf{f}_{in}^1
\end{equation}

As mentioned previously, the operation of doubling the number of layers considered is numerically stable. The inversion of matrix $\Big( \mathcal{I} - S_{22}^{(0)} S_{11}^{(n)} \Big)$ in (\ref{eq:05_03_4}) is numerically stable if the eigenvalues of $S_{22}^{(0)} S_{11}^{(n)}$ are different from unity. 
For a given eigenvector $\mathbf{x}$, $S_{11}^{(n)} \mathbf{x}$ corresponds to the fields reflected by the $2^n$ layers situated above. Multiplying the results by $S_{22}^{(0)}$ provides the fields re-reflected by the layer below. In the case of lossy layers, $S_{22}^{(0)}  S_{11}^{(n)} \mathbf{x}$ cannot reproduce $\mathbf{x}$, which in turn ensures the stability of the inversion in (\ref{eq:05_03_4}).

The convergence of the method demonstrated in Section \ref{sec:convergence} still holds for the S-matrix formalism, ensuring that accurate results without significant numerical error spreading can always be obtained with a doubly exponential convergence rate. 

%Note that using S-matrix formalism, modes propagating upward and downward have to be separated from each other. Therefore, using equivalent currents to decompose the fields, separating upward and downward modes requires extra processing. Using directly the formalism presented in Section \ref{sec:linform}, this distinction between upward and downward modes is not needed.

%the operation is unstable only if the matrix is close to be singular. On one hand, all the eigenvalues of $\mathcal{I}$ are equal to the unity. On the other hand, $S_{22}^{(i)}$ and $S_{11}^{(i)}$ describe the way fields incident on a stack of $n^i$ layers are reflected, and their spectral radius is therefore smaller or equal to the unity. The matrix $\Big( \mathcal{I} - S_{22}^{(0)} S_{11}^{(n)} \Big)$ can be singular only if the spectral

\section{Numerical results}
\label{sec:numRes}
\begin{figure}[h!]
\center
\includegraphics[width = 7.5cm]{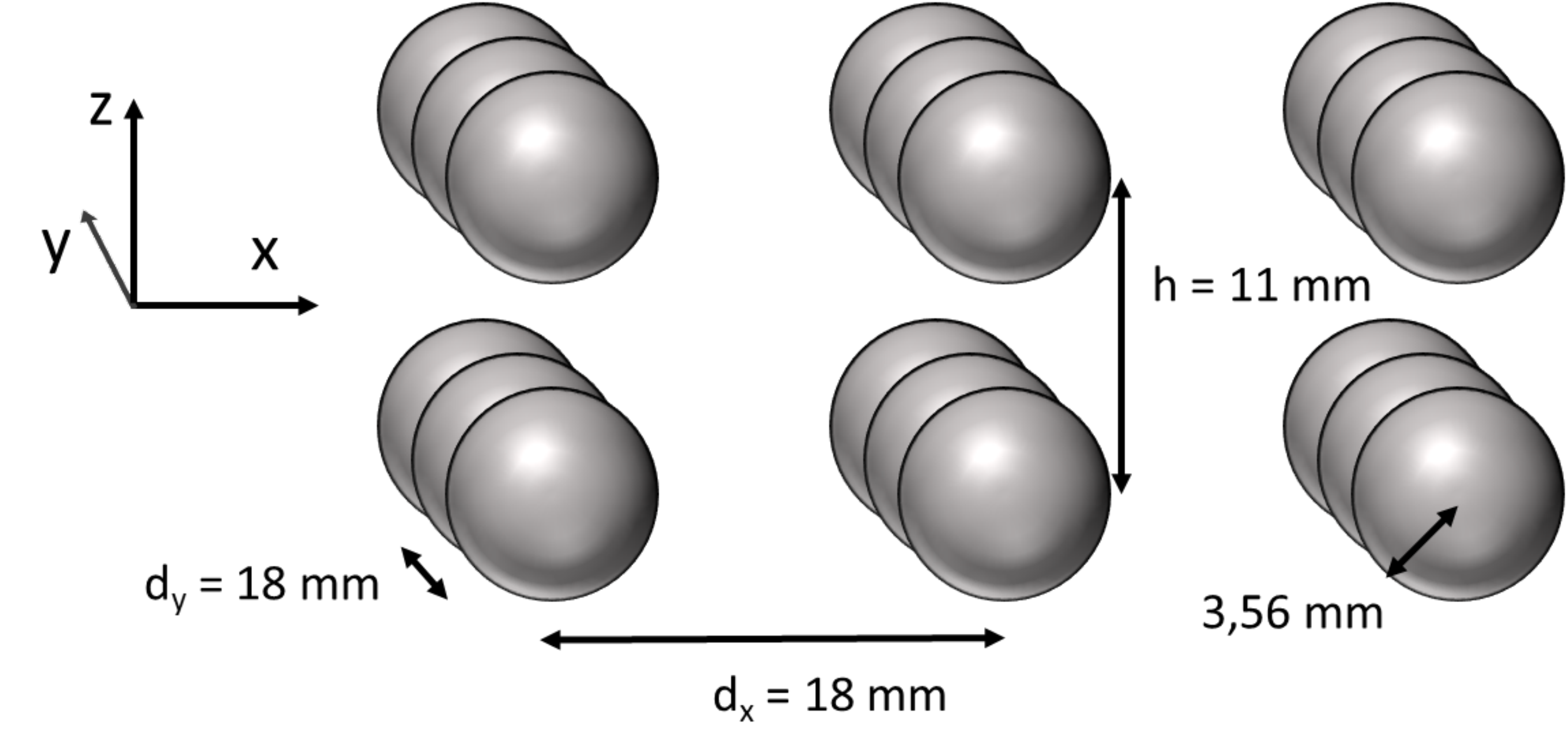}
\caption{Structure from \cite{9} used to test our method.}
\label{fig:device}
\end{figure}

To test the efficiency of the method, we considered an array of spheres, for which quasi-analytical solutions are known \cite{9}. We simulated one of the structures studied in \cite{9} (cf. Fig. \ref{fig:device}), which corresponds to a lattice of dielectric spheres of relative permittivity $\varepsilon_r = 40$ , relative permeability $\mu_r = 1$ and whose radius is 3.56 mm. Inside one layer, the spheres are arranged in a square lattice along the $\hat{x}$ and $\hat{y}$ directions with a periodicity of $d_x = d_y = 18$ mm. The periodicity along the $\hat{z}$ direction, which corresponds to the thickness of the layers constituting the metamaterial, is $h = 11$ mm. The spheres are located in vacuum.

\begin{figure}[h!]
\center
\begin{tabular}{c}
\includegraphics[width = 7cm]{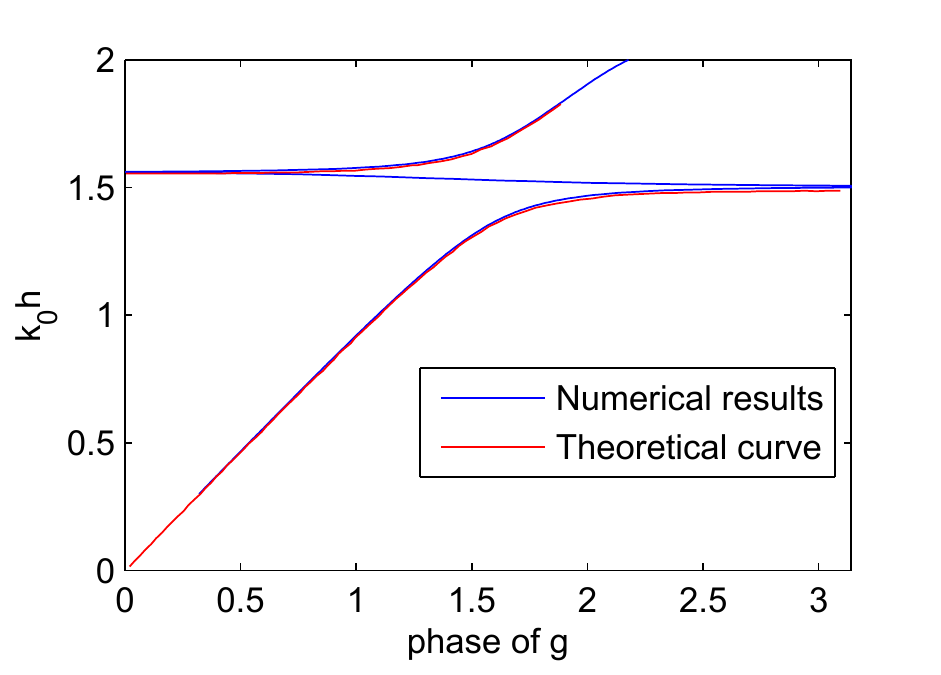}  \\ (a) \\ 
\includegraphics[width = 7cm]{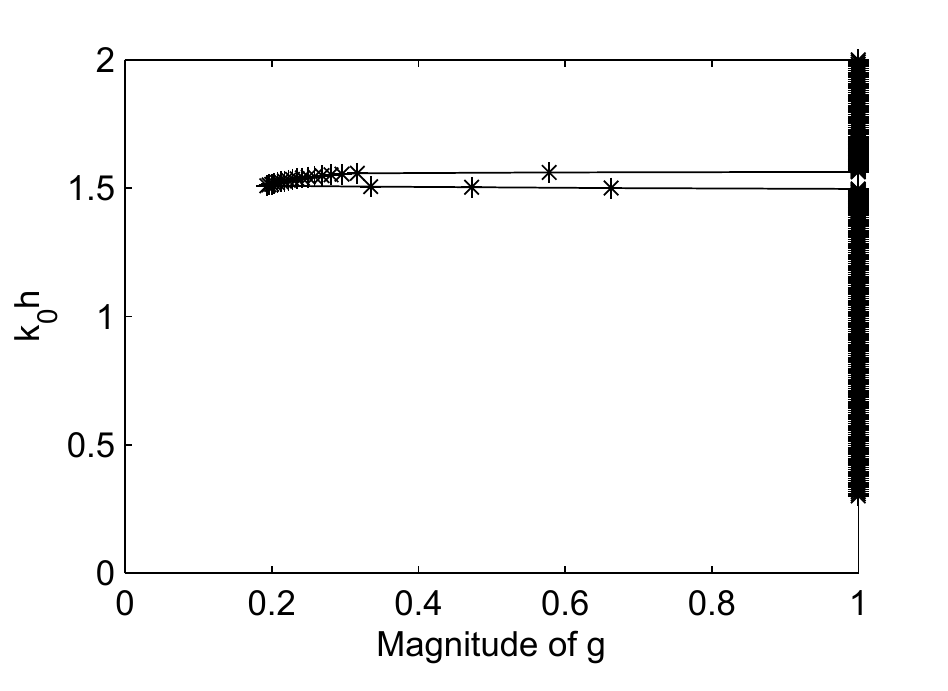} \\ (b) \\
\end{tabular}
\caption{Phase (a) and magnitude (b) of the transmission coefficient corresponding to the main propagative mode inside the structure. (a) Red: analytical results from \cite{9}, blue: numerical results.}
\label{fig:results}
\end{figure}

The eigenmodes inside the structure have been studied for several frequencies, with no phase shift between consecutive unit-cells in $\hat{x}$ and $\hat{y}$ directions inside a layer ($\varphi_x = \varphi_y = 0$). The frequency is expressed in terms of its normalized wave-vector $k_0h$, with $k_0=2\pi/\lambda_0$ and $\lambda_0$ the wavelength in free space. The phase and amplitude of the transmission coefficient $g$ corresponding to the main propagative mode can be seen in Figure \ref{fig:results}. It can be seen on graph \ref{fig:results}(b) that the transmission coefficient of evanescent modes inside the band-gap is smaller than unity.

To obtain these results, we used the linearized formulation. Losses were artificially introduced by multiplying the $\tilde{\mathcal{Z}}_{12}$ and $\tilde{\mathcal{Z}}_{21}$ matrices by a factor $(1-10^{-4})$.
The precision with which the impedance matrices have been computed being of the same order of magnitude as the loss factor, multiplication by that factor does not introduce a significant change in the accuracy of the final results. Introducing losses in that way prevents the degradation of the results near resonances, where small losses can induce huge changes in the modal distribution.

The results obtained using the linearized formulation were tested by substituting them into the parabolic formulation (\ref{eq:nonlin}) and testing the continuity of the fields at the interface between two layers. First, using the linearized formulation,
the two following quantities have been computed: (i) $\mathbf{x}_{lin}$, the equivalent currents on the interface between two consecutive layers corresponding to one mode, and (ii) $g_{lin}$, the corresponding transmission coefficient. Then, we replaced $\mathbf{x}_1$ and $\mathbf{x}_3$ in (\ref{eq:21-04_1}) by $\mathbf{x}_{lin}/g_{lin}$ and $\mathbf{x}_{lin}  ~g_{lin}$, respectively. Then, $\mathbf{x}_2$ has been computed and compared with $\mathbf{x}_{lin}$; the error $e$ is defined as

 \begin{equation}
 \label{eq:error}
 e = \dfrac{\|\mathbf{x}_{lin}-\mathbf{x}_{2}\|_2}{\|\mathbf{x}_{lin}\|_2}
 \end{equation}
 
First, using the loss factors introduced for linearization into the parabolic formulation, machine precision is achieved for all frequencies after 30 iterations. Second, the error has been checked using the non-modified impedance matrix for the parabolic formulation. The maximum error was found to be $e=10^{-4}$, which corresponds to the error introduced by the loss factor.

%We also tested the energy transported by the modes by integrating the Poynting vector over the surface of the unit-cell; the results are reported in Figure \ref{fig:poynting}. The power flow vanishes inside the band-gap, which is in accordance with the theory for lossless media.
%
%\begin{figure}[h]
%\includegraphics[width=7cm]{graphs/Poynting_2_v2.pdf}
%\caption{Power flow of the main propagative mode with respect to frequency. As expected, no energy is carried by the mode inside the band-gap (dotted lines).}
%\label{fig:poynting}
%\end{figure}

\begin{figure}[h!]
\center
\begin{tabular}{c}
\includegraphics[width = 7.5cm]{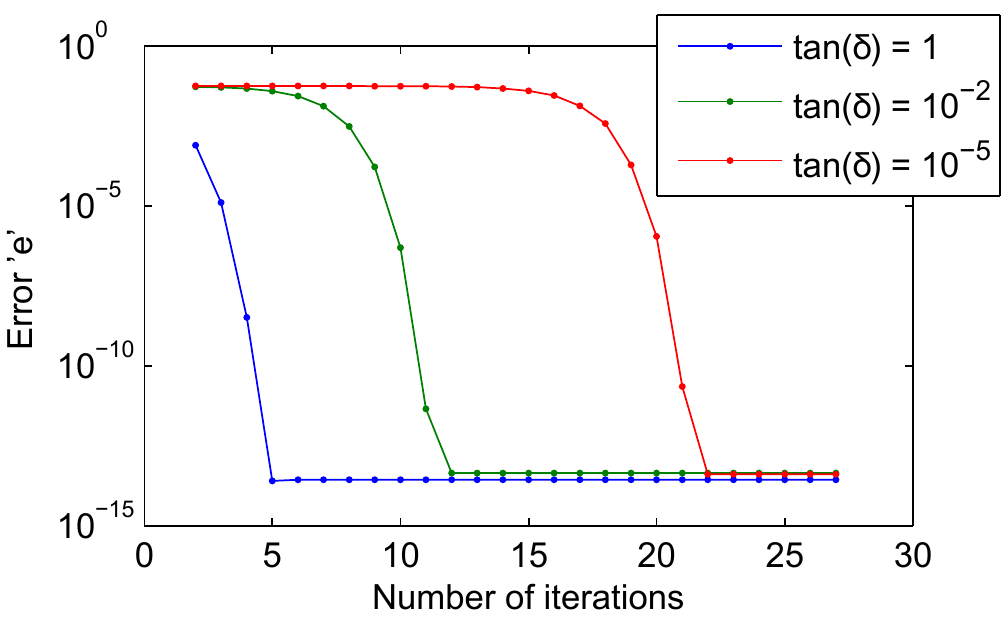}  \\ (a) \\ 
\includegraphics[width = 7.5cm]{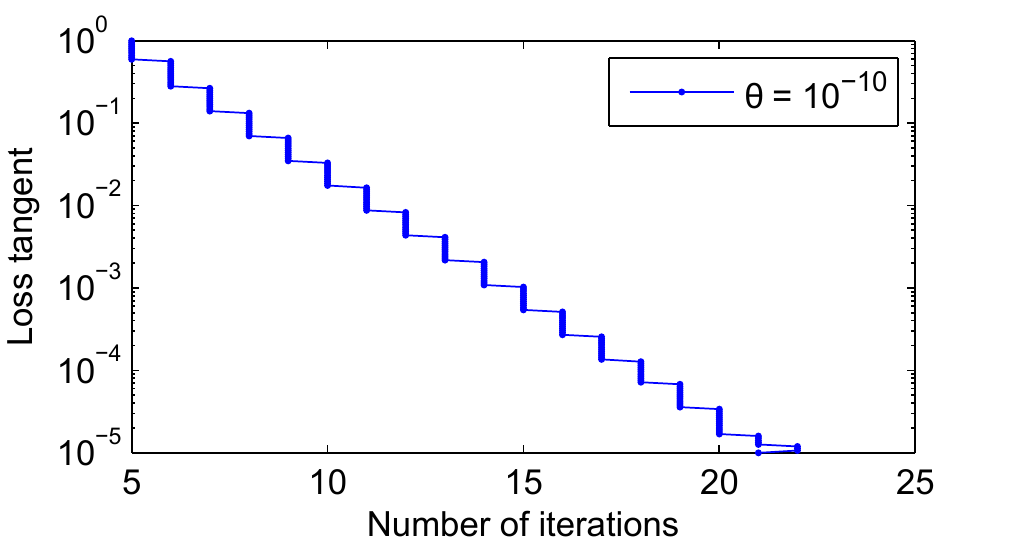} \\ (b) \\
\end{tabular}
\caption{Convergence of the method. (a) Evolution of the error with the number of iterations for different amount of losses. (b) Relation between the losses and the number of iterations required to reach an error $e$ smaller than a threshold $e_0 =10^{-10}$.   }
\label{fig:convergence}
\end{figure}

The convergence of the method has also been tested when introducing a varying amount of conduction losses inside the metamaterial. This has been implemented adding a complex component $\varepsilon''$ to the permittivities of the air and of the dielectric. The loss tangent $\tan(\delta) = \varepsilon''/\varepsilon$ has been made identical for air and dielectric, the frequency has arbitrarily been chosen to be such that $k_0h = 1$. Results are shown in Figure \ref{fig:convergence}. On the top graph, we can see the doubly exponential convergence of the method for a fixed amount of losses (cf. Equation (\ref{eq:convergence})).  On the second graph, 
the number of iterations required to reach an error $e$ chosen smaller than a predefined threshold $e_0 = 10^{-10}$ is shown. The predicted exponential convergence with respect to the losses is clearly visible.

\begin{figure}[h!]
\center
\includegraphics[width=7.5cm]{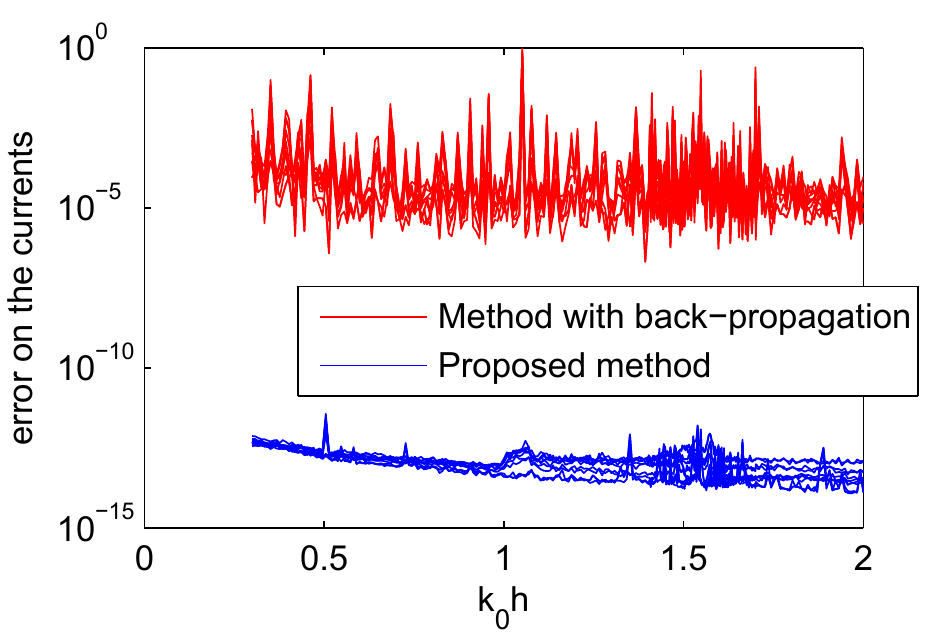}
\caption{Numerical error introduced when computing the first ten eigenmodes of the structure using two different linear methods. Red curves: linear formulation that has to handle back-propagation (cf. Section \ref{sec:non-lin}). Blue curves: proposed method.}
\label{fig:errbackpropag}
\end{figure}

We also compared the accuracy of the results that can be achieved using the proposed method with the accuracy achieved by linear methods handling back-propagation. Using the same set of matrices as the one used to provide the results displayed on Fig. \ref{fig:results}, we computed the eigenmodes of the structure using the proposed formulation and the one based on doubling the number of unknowns, which is provided at the end of Section \ref{sec:non-lin} (Equation (\ref{eq:17-05-1})). The numerical error introduced by both techniques for the first ten eigenmodes was computed using (\ref{eq:error}) and is shown in Fig. \ref{fig:errbackpropag}. The error is only originating from the numerical round-off error. 

First, we can notice that numerical error is limited when we use the method we presented (blue curves), due to the fact that every step of the method is well-conditioned. The increase in the error for small wave vectors $k_0 h$ is probably due to the ``low frequency breakdown" that is related to the MoM simulation technique rather than on our method for determining eigenmodes.

Second, as explained in \cite{limitations}, methods that have to handle back-propagation are ill-conditioned. Therefore, the numerical noise tends to grow during the computation of the modes (red curves). 

  As we can see, the accuracy of the proposed method is between five and ten orders of magnitude better than the accuracy provided by techniques handling back-propagation.

We also compared the accuracy of the proposed method using S-matrix formalism with the accuracy that can be achieved when using the associated T-matrix.
The S-matrix of the layer has been obtained from the MoM simulation results. One layer of metamaterial has been excited by several propagative and evanescent plane waves, so that the total number of degrees of freedom is preserved, i.e. we used as many plane waves as we used basis functions to describe the currents on the interface between consecutive layers. 
 
The error has been computed as follows. First, the eigenmodes and their associated eigenvectors have been computed using the methods presented in Section \ref{sec:Smat}. The eigenfields associated to these eigenvectors can be decomposed into fields incident on the layer $\mathbf{f}^1_{in}$ and $\mathbf{f}^2_{in}= g \mathbf{f}^1_{out}$, and fields scattered by the layer $\mathbf{f}^1_{out}$ and $\mathbf{f}^2_{out} = g \mathbf{f}^1_{in}$. The incident fields were multiplied by the S-matrix of one layer, and the resulting vector were compared with the expected scattered fields from Equations (\ref{eq:01-12-01}).

\begin{figure}[h!]
\center
\includegraphics[width=7.5cm]{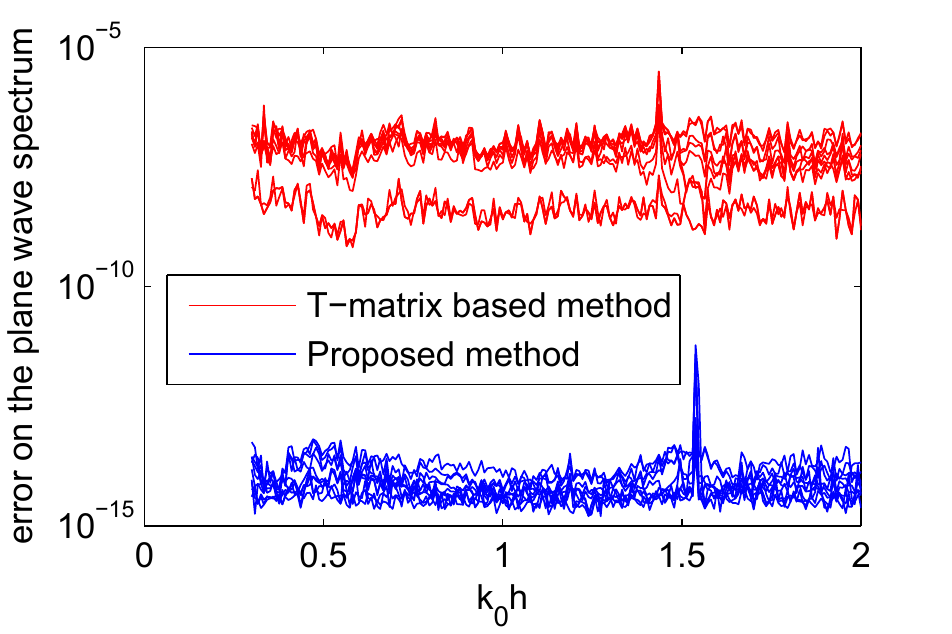}
\caption{Numerical error for the computation of the first ten eigenmodes using the S-matrix of the structure. Red curves: eigenvalue decomposition of the T-matrix. Blue curves: iterative method.}
\label{fig:errsmat}
\end{figure}

 The results can be seen in Fig. \ref{fig:errsmat}. The numerical error has been computed for the first ten eigenmodes of the structure. The red curves represent the error obtained when the T-matrix of the layer is computed from its S-matrix, while the blue curves represent the error obtained using the iterative method presented in Section \ref{sec:Smat}. 
 
First, we can notice, as mentioned previously, that using a plane wave decomposition of the fields tends to improve the accuracy reached due to natural filtering of the ill-conditioned modes. Indeed, the error of the red curves in Fig. \ref{fig:errsmat} is much lower than the error in Fig. \ref{fig:errbackpropag}. Moreover, the lower red curves of Fig. \ref{fig:errsmat} are corresponding to the less evanescent eigenmodes of the structure. They are characterized by a lower contribution of the highly evanescent plane waves, which explain the better filtering, thus the lower numerical noise. 

Second, we can see that the error obtained when stacking the layers remains close to the machine precision. This is a direct consequence of the fact that all the matrix operations involved in computing the eigenmodes of the structure are well-conditioned. 

Finally, the time for filling the impedance matrix for a fixed frequency and transverse phase shift ranged from 1h55' to 2h40' depending on frequency, while the time spent to compute all the eigenmodes and eigenvalues after 30 iterations for that frequency and phase shift took less than 20 seconds, which is negligible with respect to matrix filling.

\section{Conclusion}
We presented a method based on the Method of Moments to compute the eigenmodes inside 3D periodic metamaterials. The modes are characterized by a field distribution at the interface between consecutive layers and a complex transmission coefficient of the fields from one layer to the next. First, we presented a parabolic formulation, which is based on the continuity of tangential fields across consecutive layers. Then, we proposed an iterative technique that can be applied to the simulation results to linearize the problem, so that all the modes and the corresponding transmission coefficients can be found using a single eigenvalue decomposition of a matrix. 

We also extended the proposed method to the more general Scattering-matrix formalism and compared it with other methods based on Transfer-matrix. We showed how the computation of the T-matrix, which can be numerically unstable, is avoided in our case to provide a fully stable method.

The linearization method, which does not entail any approximation, is based on the stacking of layers. At each iteration, the number of stacked layers is doubled. Adding arbitrarily small losses, a doubly exponential convergence toward the solution of the parabolic formulation is demonstrated. We showed that, to reach a given accuracy, the time devoted to the iterative technique increases linearly for exponentially decreasing losses inside the material.

%%%%%%%%%%%%%%%%%%%%%%%%%%%%%%%

\begin{appendices}

\section{From M\"{u}ller to T-matrix}
\label{sec:muller}
In this section, we will draw the link between the T-matrix based method described in \cite{4} and the eigenmodes determination problem based on the M\"{u}ller formulation for the Method of Moments\cite{Muller}. 

Using the PMCHWT formulation of the MoM, the equivalent currents on the interfaces between different media are found by imposing the continuity of the tangential electric and magnetic fields on both sides of the testing functions. Using the M\"{u}ller formulation, however, the equivalent currents along the interfaces separating different media are found imposing on both sides of the interface:
%Using the M\"{u}ller formulation, the equivalent currents on interfaces separating different media are not found by using the continuity of the fields, but by directly relating the local current density to the local value of the total electric and magnetic fields.

\begin{subequations}
\label{eq:A1}
\begin{align}
\hat{n} \times \mathbf{E}^{in} + \hat{n} \times \mathbf{E}^{sc} = -\mathbf{M} \\
\hat{n} \times \mathbf{H}^{in} + \hat{n} \times \mathbf{H}^{sc} = \mathbf{J}
\end{align}
\end{subequations}

with $\hat{n}$, the direction normal to the surface toward the side of the interface that is tested, $ \mathbf{E}^{in}$ and $\mathbf{H}^{in}$ the incident electric and magnetic fields and $\mathbf{E}^{sc}$ and $\mathbf{H}^{sc}$ the electric and magnetic fields radiated on the interface by $\mathbf{J}$ and $\mathbf{M}$, the unknown equivalent electric and magnetic currents. 

To apply M\"{u}ller formulation to the geometry illustrated in Fig. \ref{fig:1layer}, it is necessary to impose (\ref{eq:A1}) on both side of both interfaces.

Below, we concentrate on the inner part of the layer. The surfaces are discretized and (\ref{eq:A1}) is applied to both interfaces on their inner side. The system of equations to solve then reads

\begin{equation}
\label{eq:A2}
\tilde{Z}_{M,11}^{\uparrow} \mathbf{x}_1 - \tilde{Z}_{M,12} \mathbf{x}_2 = \mathbf{x}_1
\end{equation}
\begin{equation}
\label{eq:A3}
-\tilde{Z}_{M,22}^{\downarrow} \mathbf{x}_2 + \tilde{Z}_{M,21} \mathbf{x}_1 = -\mathbf{x}_2
\end{equation}

with $\tilde{Z}_{M,pq}$ the impedance matrix specific to the M\"{u}ller formulation. This impedance matrix describes the transverse fields generated on Plane $p$ by currents on Plane $q$ using the periodic Green's function, and the $\uparrow$ and $\downarrow$ symbols indicate if fields are tested just above or just below the interface. 

As well known, the use of the equivalence principle produces twice as many equations as unknowns. The M\"{u}ller formulation can be chosen to avoid a non-linear eigenmode formulation by involving only two consecutive interfaces.

%It can seem that each system of equations can be solved, as there are as many equations as unknowns. However this is not the case as every equation of the system of equations is not linearly independent from all the others. A simple illustration of the remaining degrees of freedom is that the solution has to depend also on the sources situated outside the layer. 

In that case, half of the unknowns will drop using Equation (\ref{eq:periodic}).
We then obtain

\begin{equation}
\tilde{Z}_{M,12}^{-1} \big( \tilde{Z}_{M,11}^\uparrow - \mathcal{I} \big) \mathbf{x}_1 = g \mathbf{x}_1
\end{equation}
\begin{equation}
\big( \tilde{Z}_{M,22}^\downarrow - \mathcal{I} )^{-1} \tilde{Z}_{M,21} \mathbf{x}_1 = g \mathbf{x}_1
\end{equation}

As always when using M\"{u}ller formulation, these two system of equations have to be combined somehow \cite{Muller}. One possibility is to simply add them together. However, to keep the eigenvalue $g$ constant, we will define the T-matrix as

\begin{equation}
\label{eq:A4}
\mathcal{T} = \dfrac{\tilde{Z}_{M,12}^{-1} \big( \tilde{Z}_{M,11}^\uparrow - \mathcal{I} \big) + \big( \tilde{Z}_{M,22}^\downarrow - \mathcal{I} )^{-1} \tilde{Z}_{M,21}}{2}
\end{equation}

such that

\begin{equation}
\mathcal{T} \mathbf{x}_1 = g \mathbf{x}_1
\end{equation}

We can notice, as mentioned previously, that the T-matrix formulation has to deal with back-propagation, which is illustrated by term $\tilde{Z}_{M,12}^{-1} $ in formula (\ref{eq:A4}).

With this definition, the eigenmodes inside the structure correspond to the eigenvectors and eigenvalues of $\mathcal{T}$ and can be computed all at once. However, because of the back-propagation term, $\mathcal{T}$ is ill-conditioned and the results may be inaccurate.

\section{Decay of modes due to loss}
\label{app:B}
A more formal proof of Equation (\ref{eq:28-05-08}) is presented here, based on energy conservation. The physical origin of the constant $\rho_\alpha$ is also investigated.

%Consider a semi-infinite metamaterial made of a semi-infinite superposition of layers. On the interface between this metamaterial and the background medium, a mode $\alpha$ is excited. Due to the fact that this mode is, by definition, a homogeneous solution to Maxwell's equations for the geometry of the metamaterial, it can only couple with other modes through defects in this geometry's {\color{red} periodicity}. The metamaterial being infinite in the direction of propagation of the mode (i.e. the direction of its energy flow), the latter will never meet such a defect in the periodicity and therefore scattering into other modes cannot happen.

Consider an infinitely 3D periodic metamaterial made of an infinite superposition of layers. A mode $\alpha$ is excited at the bottom interface of a layer, say Plane 0 and Layer 0, and propagates upward. Due to the fact that this mode is, by definition, a homogeneous solution to Maxwell's equations for the periodic geometry of the metamaterial, this mode will propagate without coupling with other modes, unless there is some defect in the periodic geometry. As we are considering an infinite metamaterial, such defect does not exist and therefore the modes $\alpha$ can be treated independently of the other modes, with which it cannot interact. In the following, we will consider that the direction of propagation of a mode is defined as the direction of its energy flow. For evanescent modes, which do not carry energy, the direction is given by the energy flow transported by the mode if infinitesimal losses are introduced.

This mode $\alpha$ is characterized by the fields distribution $(\mathbf{E}_\alpha(\mathbf{r}), \mathbf{H}_\alpha(\mathbf{r}))$ and a transfer coefficient $g_\alpha$. The power transported by that mode crossing interface $p$, between layers $p-1$ and $p$, is given by

\begin{equation}
P_\alpha^{p} = \Re\Big\{ \iint_{S_p} \dfrac{\mathbf{E}_\alpha(\mathbf{r}) \times \mathbf{H}_\alpha^*(\mathbf{r})}{2} \cdot \hat{n} ~ dS \Big\}
\end{equation}

with $\Re\{A\}$ the real part of $A$, $\hat{n}$ the normal to the interface $p$ toward layer $p$ and $S_p$ the surface of that interface limited to one unit cell of the metamaterial. Using Equation (\ref{eq:periodic}) provides:
\begin{equation}
\label{eq:01-06-01}
P_\alpha^{p} = |g_\alpha|^{2p} \Re\Big\{  \iint_{S_0} \dfrac{\mathbf{E}_\alpha(\mathbf{r}) \times \mathbf{H}_\alpha^*(\mathbf{r})}{2} \cdot \hat{n} ~ dS \Big\}
\end{equation}

with the integration that is now performed over $S_0$. The medium being lossy, the power crossing two consecutive interfaces has to fade accordingly to the losses, i.e:
\begin{equation}
\label{eq:01-06-03}
P_\alpha^{p} - P_{\alpha}^{p+1} = P_{\alpha,loss}^{p} > 0
\end{equation}

with $P_{\alpha,loss}^{p}$ the losses due to mode $\alpha$ inside layer $p$. Using Equation (\ref{eq:01-06-01}) and noticing that the power crossing the interface has to be positive due to causality, one can find

\begin{equation}
1-|g_\alpha|^2 > 0
\end{equation}

which corresponds to the physical constraint that the amplitude of the mode has to decrease from layer to layer ($|g_\alpha|<1$).

Now, consider that losses are introduced using a non-zero conductivity $\sigma$. The amount of losses due to mode $\alpha$ inside layer $p$, $P_{\alpha ,loss}^p$, can be computed using:

\begin{equation}
P_{\alpha,loss}^{p} = \iiint_{V_p} \sigma \dfrac{{\|\mathbf{E}_\alpha\|_2}^2}{2} dV
\end{equation}

with $\sigma$ the local conductivity and $V_p$ the volume corresponding to the intersection between the layer $p$ and one unit cell of the metamaterial. Note that $\mathbf{E}$ depends on the value of $\sigma$. Indeed, changing the value of $\sigma$ modifies the Green's function of the lossy medium and therefore the solution to the scattering problem. However, if the losses are small enough, the change in the shape of the fields for decreasing losses becomes negligible and a perturbation based approach can be used. If we consider that $\sigma$ is small and constant in the whole lossy parts of the metamaterial, we can introduce the constant $D_\alpha$ that is independent of the conductivity such that

\begin{equation}
D_\alpha \equiv \iiint_{V_0^{loss}} \dfrac{{\|\mathbf{E}_\alpha|_{\sigma = 0}\|_2}^2}{2} dV
\end{equation}

with $V_0^{loss}$, the volume corresponding to the lossy part of the first layer limited to one unit cell of the metamaterial.
Then the power loss will asymptotically reach

\begin{equation}
\lim_{\sigma \rightarrow 0} \dfrac{P_{\alpha,loss}^{p}}{\sigma} = |g_\alpha|^{2p} D_\alpha
\end{equation}

Combining this expression with (\ref{eq:01-06-01}) and (\ref{eq:01-06-03}), it provides:

\begin{equation}
\label{eq:24-02-16-06}
|g_\alpha|^2 \simeq 1- 2\rho_\alpha \sigma
\end{equation}
with 
\begin{equation}
\rho_\alpha = \dfrac{D_\alpha}{ \Re\Big\{ \iint_{S_0} (\mathbf{E}_\alpha \times \mathbf{H}_\alpha^*) \cdot \hat{n} ~ dS \Big\}}
\end{equation}
which is proportional to the ratio between the power dissipated and the power transported by one mode.

If the conductivity is small enough, from (\ref{eq:24-02-16-06}), the transmission coefficient can be expressed using a first-order Taylor series:
\begin{equation}
|g_\alpha| \simeq 1-\rho_\alpha \sigma
\end{equation}

\end{appendices}

\section*{Acknowledgments}

Parts of this work has been supported by the Research Networking Program NEWFOCUS of the European Science Foundation "New Frontiers in mm/sub-mm waves integrated dielectric focusing systems".

This research has been partially funded by the Interuniversity Attraction Poles Programme 7/23 BESTCOM initiated by the Belgian Science Policy Office. 

Computational resources have been provided by the supercomputing facilities of the Universit\'{e} catholique de Louvain (CISM/UCL) and the Consortium des Equipements de Calcul Intensif en F\'{e}d\'{e}ration Wallonie Bruxelles (CECI) funded by the Fond de la Recherche Scientifique de Belgique (F.R.S.-FNRS) under convention 2.5020.11.

The authors would like to thank the reviewers for their fruitful comments.

\begin{IEEEbiography}[{\includegraphics[width=1in,height=1.25in,clip,keepaspectratio]{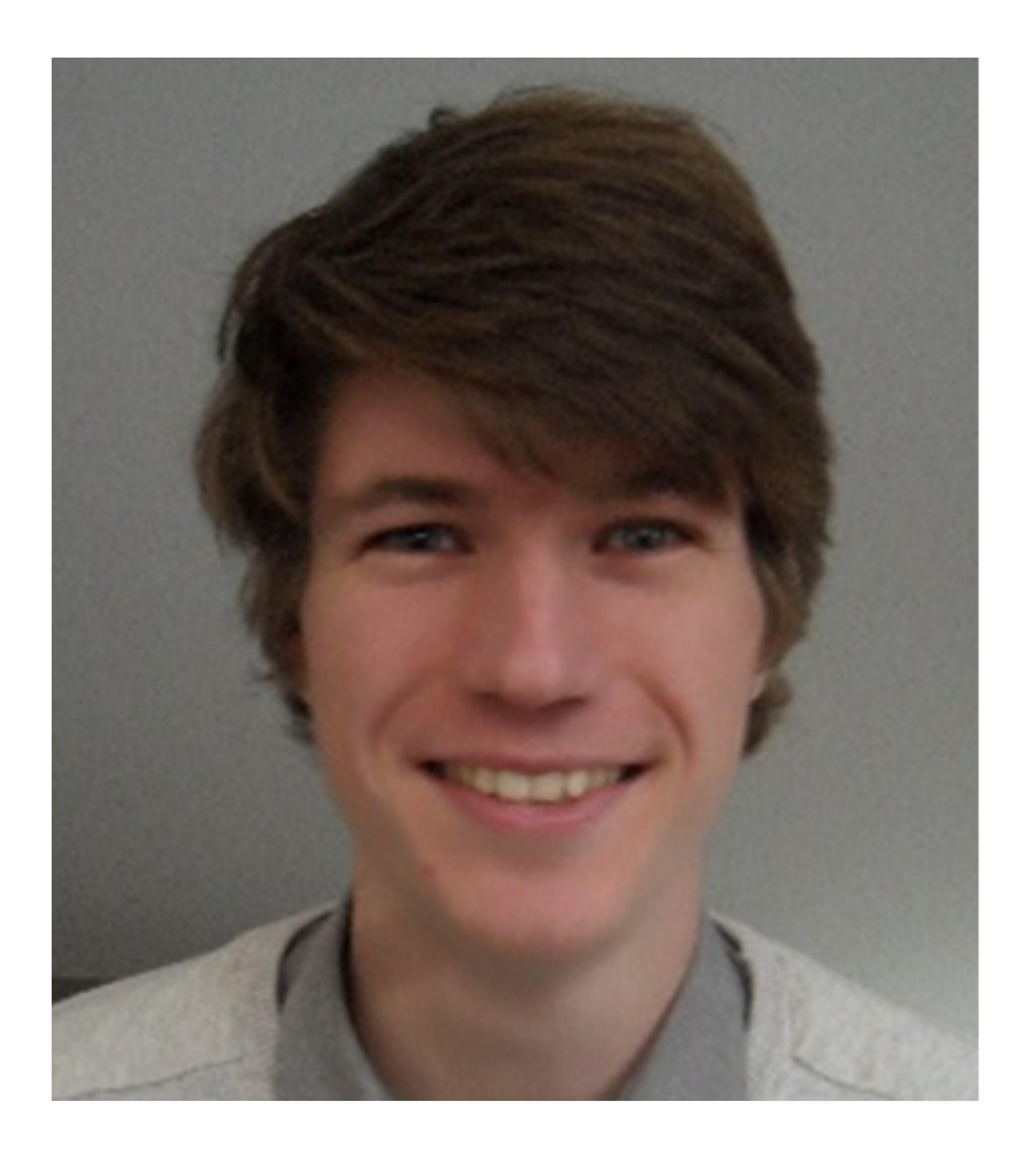}}]{Denis Tihon}
received the B. Sc. and the M. Sc. degrees in Universit\'{e} catholique de Louvain (UCL), Louvain-la-Neuve, Belgium, in 2011 and 2013 respectively. Since 2013, he is working as a Teaching Assistant and pursuing a PhD Thesis in the institute of Information and Communication Technologies, Electronics and Applied Mathematics (ICTEAM) in UCL. He is also studying the absorption of partially coherent fields in collaboration with the Cavendish Laboratory, Cambridge, U.K.
His domains of interests include the modeling of metamaterials, the integral equation based methods and the absorption of partially coherent electromagnetic fields.
\end{IEEEbiography}

\begin{IEEEbiography}[{\includegraphics[width=1in,height=1.25in,clip,keepaspectratio]{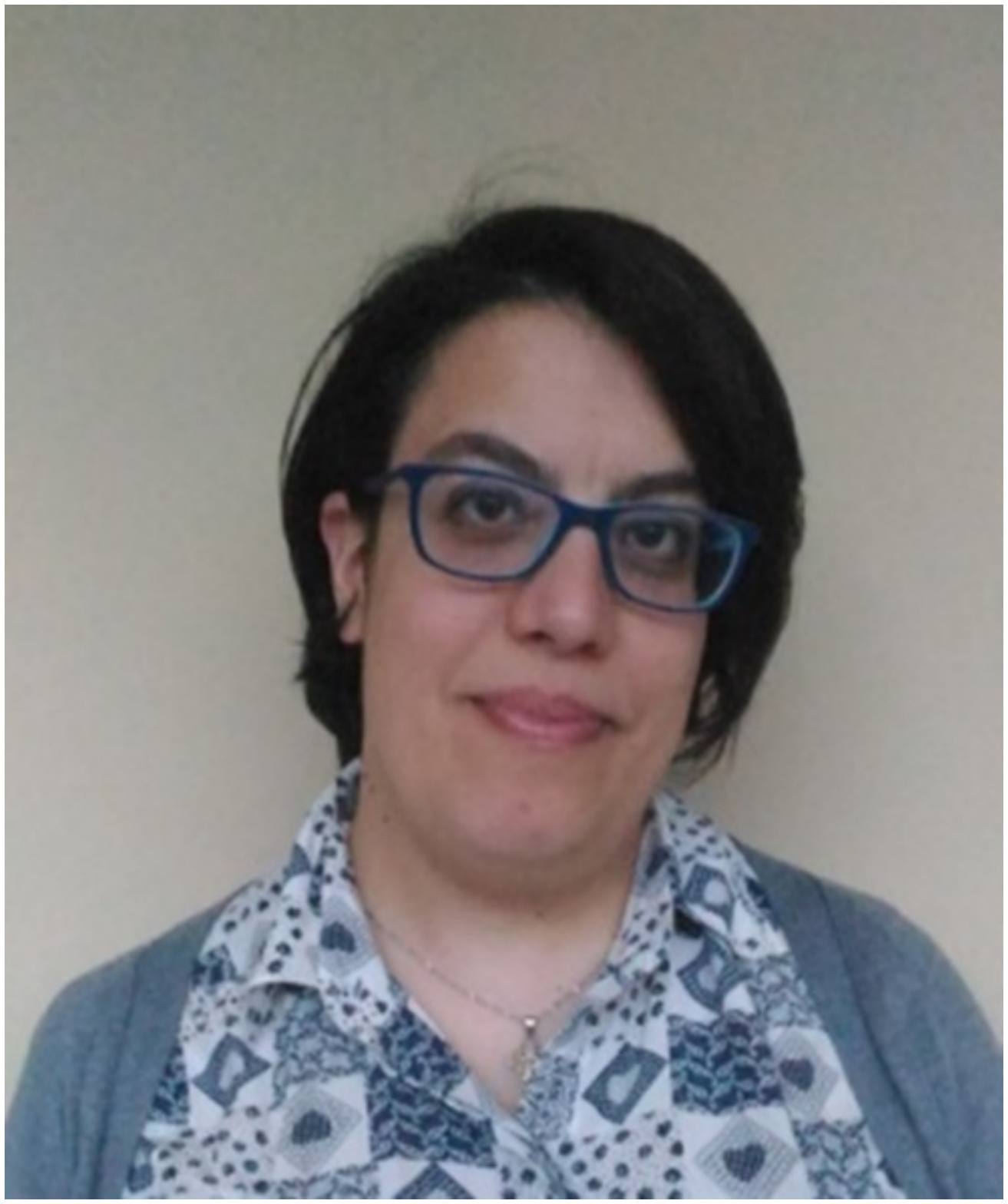}}]{Valentina Sozio}
received the B. Sc. and the M. Sc. in Telecommunications Engineering at the Sapienza University of Rome. She obtained the PhD in Information Engineering at the University of Siena in 2015 with research on homogenization techniques for 3D metamaterials. In 2014 she spent six months at the Universit\'{e} catholique de Louvain, where she investigated numerical methods for metamaterials characterization. Since 2015 she has been working at the Istituto Superiore Mario Boella in Turin, as a researcher within the Antenna and EMC Unit.
\end{IEEEbiography}

\begin{IEEEbiography}[{\includegraphics[width=1in,height=1.25in,clip,keepaspectratio]{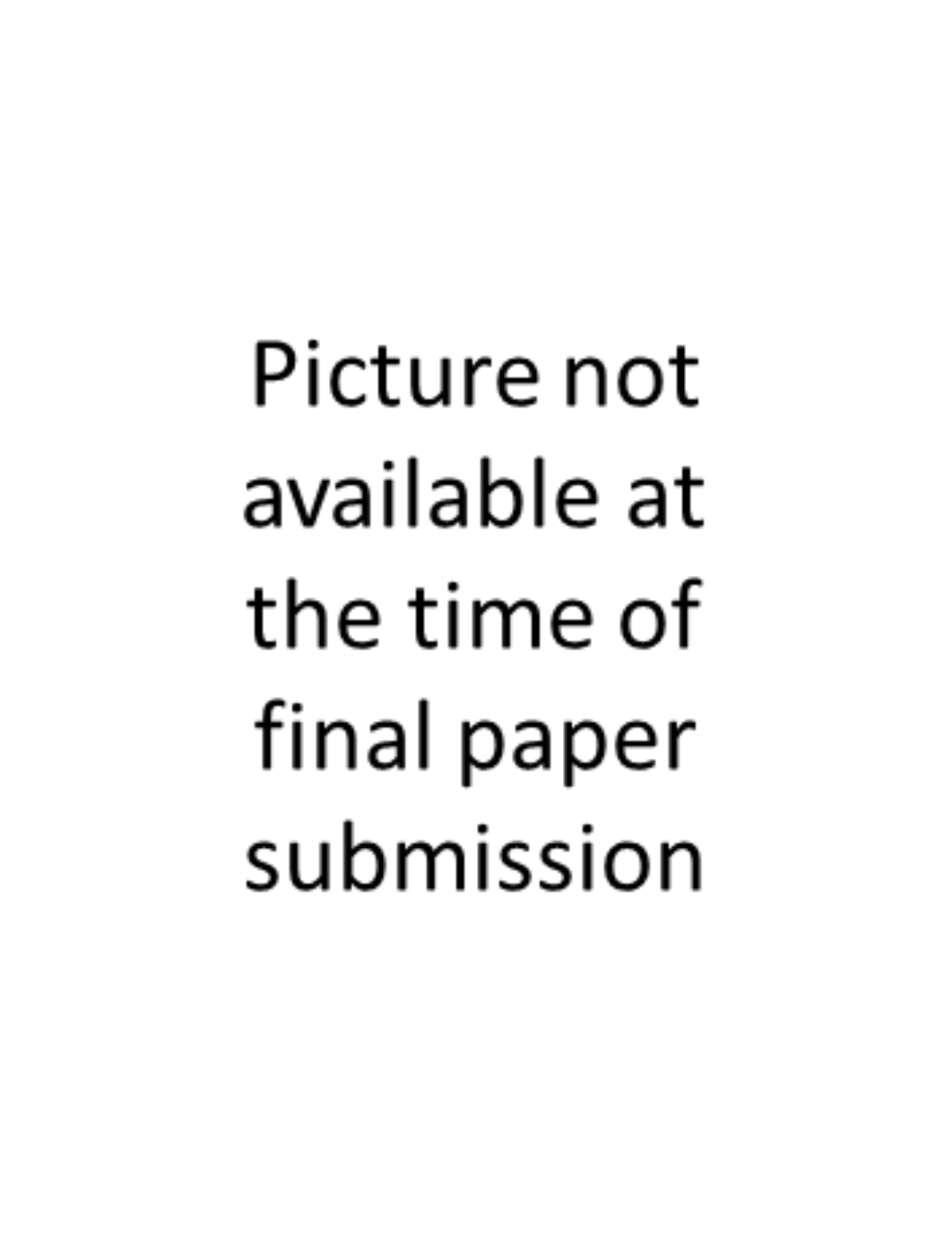}}]
{Nilufer Aslihan Ozdemir} (M'07) received the B. Sc. and the M. Sc. degrees in electrical and electronics engineering from the Middle East Technical University, Ankara, Turkey, in 1997 and 2000, respectively, and the Ph.D. degree in electrical and computer engineering from The Ohio State University, Columbus, in 2007.
In 2007, She worked in the Universit\'{e} catholique de Louvain, Louvain-la-Neuve, Belgium, as a Postdoctoral Researcher. Her research has focused on integral equation based numerical solution of electromagnetic scattering and radiation problems. 
Since 2015, Dr. Ozdmir is active with the Royal Observatory of Belgium.
\end{IEEEbiography}

\begin{IEEEbiography}[{\includegraphics[width=1in,height=1.25in,clip,keepaspectratio]{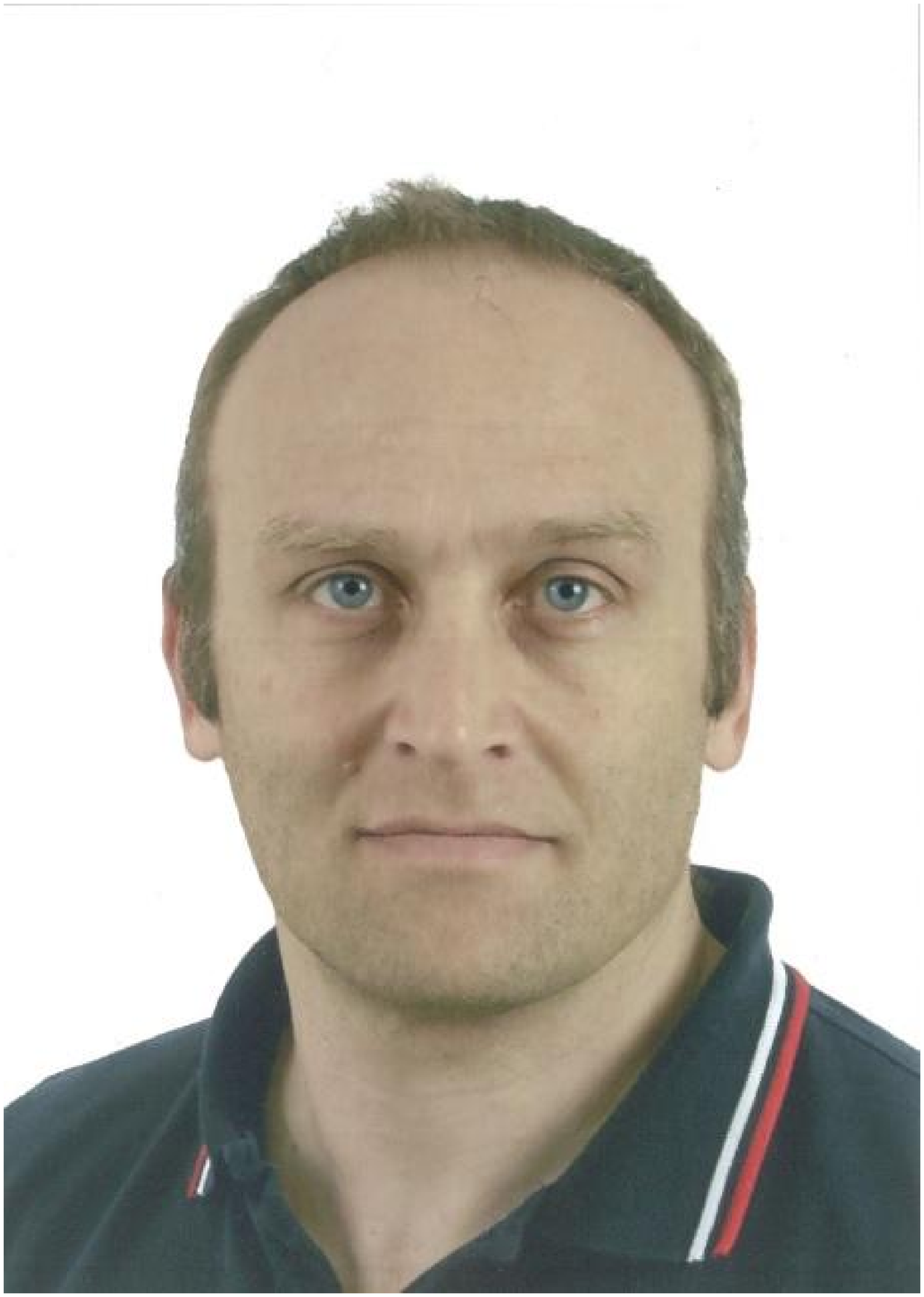}}]{Matteo Albani} (M'98-SM'10) received the Laurea degree in electronic engineering and Ph.D. degree in telecommunications engineering from the University of Florence, Florence, Italy, in 1994 and 1999, respectively.
He is currently an Associate Professor with the Information Engineering and Mathematics Department, University of Siena, Siena, Italy, where he is also Director of the Applied Electromagnetics Laboratory. From 2001 to 2005, he was an Assistant Professor with the University of Messina, Messina, Italy.
He coauthored more than 65 journal papers and more than 170 conference papers. His research interests are in the areas of high-frequency methods for electromagnetic scattering and propagation, numerical methods for array antennas, antenna analysis and design, metamaterials.
Dr. Albani received the ``G. Barzilai" Young Researcher Best Paper Award, at the XIV RiNEM, Ancona, Italy 2002, and the URSI Commission B Young Scientist Award at 2004 URSI EMTS, Pisa, Italy. He was Co-author and Advisor of the winners of the Best Paper Award, at the First European AMTA Symp. 2006, Munich, Germany and of the ``3rd Prize Young Scientist Best Paper Award" 2010 URSI EMTS, Berlin, Germany. With his co-authors. he was awarded with the Antenna Theory Best Paper at the EuCAP 2014 in den Haag, the Netherlands.
He is a member of EURAAP and URSI.

\end{IEEEbiography}

\begin{IEEEbiography}[{\includegraphics[width=1in,height=1.25in,clip,keepaspectratio]{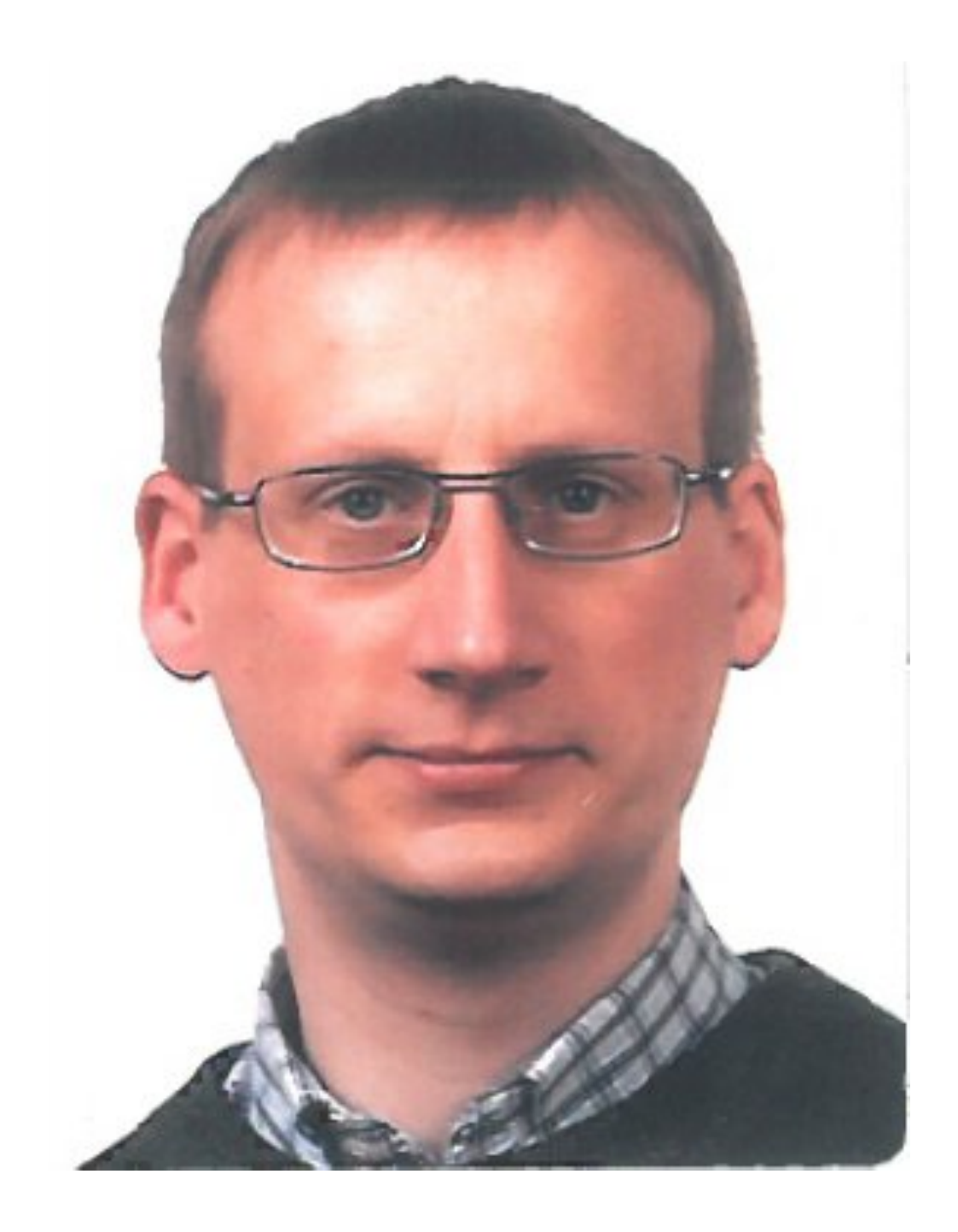}}]{Christophe Craeye} (M'98-SM'11) was born in
Belgium in 1971. He received the Electrical
Engineering and Bachelor in Philosophy degrees
and the Ph.D. degree in applied sciences from the
Universit\'{e} catholique de Louvain (UCL), Louvain-la-Neuve,
Belgium, in 1994 and 1998, respectively.
From 1994 to 1999, he was a Teaching Assistant
with UCL and carried out research on the radar
signature of the sea surface perturbed by rain, in collaboration
with NASA and ESA. From 1999 to 2001,
he was a Postdoctoral Researcher with the Eindhoven
University of Technology, Eindhoven, The Netherlands. His research there was
related to wideband phased arrays devoted to the square kilometer array radio
telescope. In this framework, he was also with the University of Massachusetts,
Amherst, MA, USA, in the Fall of 1999, and was with the Netherlands Institute
for Research in Astronomy, Dwingeloo, The Netherlands, in 2001. In 2002,
he started an antenna research activity at the Universit\'{e} catholique de Louvain,
where he is now a Professor. He was with the Astrophysics and Detectors
Group, University of Cambridge, Cambridge, U.K., from January to August
2011. His research is funded by R\'{e}gion Wallonne, European Commission, ESA,
FNRS, and UCL. His research interests include finite antenna arrays, wideband
antennas, small antennas, metamaterials, and numerical methods for fields in
periodic media, with applications to communication and sensing systems.
Prof. Craeye was an Associate Editor of the \textit{IEEE transactions on antennas and propagation} from 2004 to 2010. He currently serves as an Associate
Editor of the \textit{IEEE Antennas and Wireless Propagation Letters}. In 2009, he was
the recipient of the 2005-2008 Georges Vanderlinden Prize from the Belgian
Royal Academy of Sciences.

\end{IEEEbiography}

\end{document}